\documentclass[aps,pra,twocolumn,superscriptaddress,amsmath,amssymb,showpacs,,]{revtex4-1}

\usepackage{graphicx}	
\usepackage{dcolumn}	
\usepackage{bm}		
\usepackage{verbatim} 	
\usepackage{braket}
\usepackage{xr}
\usepackage{hyperref}
\usepackage{booktabs}
\usepackage{upgreek}
\usepackage{notes2bib}
\usepackage{pdfpages}

\makeatletter
\AtBeginDocument{\let\LS@rot\@undefined}
\makeatother

\newcolumntype{L}[1]{>{\raggedright\let\newline\\\arraybackslash\hspace{0pt}}m{#1}}
\newcolumntype{C}[1]{>{\centering\let\newline\\\arraybackslash\hspace{0pt}}m{#1}}
\newcolumntype{R}[1]{>{\raggedleft\let\newline\\\arraybackslash\hspace{0pt}}m{#1}}

\newcommand{\resim}{\mathord{\sim}}

\begin{document}

\title{Single Photon Detection by
Cavity-Assisted All-Optical Gain}

\author{Christopher Panuski}
\email{cpanuski@mit.edu}
\affiliation{Department of Electrical Engineering and Computer Science, MIT, Cambridge, MA 02139, USA}

\author{Mihir Pant}
\affiliation{Department of Electrical Engineering and Computer Science, MIT, Cambridge, MA 02139, USA}

\author{Mikkel Heuck}
\affiliation{Department of Photonics Engineering, Technical University of Denmark, 2800 Kgs. Lyngby, Denmark}

\author{Ryan Hamerly}
\affiliation{Department of Electrical Engineering and Computer Science, MIT, Cambridge, MA 02139, USA}

\author{Dirk Englund}
\email{englund@mit.edu}
\affiliation{Department of Electrical Engineering and Computer Science, MIT, Cambridge, MA 02139, USA}

\begin{abstract}
We consider the free carrier dispersion effect in a semiconductor nanocavity in the limit of discrete photoexcited electron-hole pairs. This analysis reveals the possibility of ultrafast, incoherent transduction and gain from a single photon signal to a strong coherent probe field. Homodyne detection of the displaced probe field enables a new method for room temperature, photon-number-resolving single photon detection. In particular, we estimate that a single photon absorbed within a silicon nanocavity can, within tens of picoseconds, be detected with $\resim99\%$ efficiency and a dark count rate on the order of kHz assuming a mode volume $V_\text{eff}\sim 10^{-2}(\lambda/n_\text{Si})^3$ for a 4.5 $\upmu$m probe wavelength and a loaded quality factor $Q$ on the order of $10^4$. 
\end{abstract}

\maketitle

\section{Introduction}
\vspace{-10pt}
An outstanding goal in optoelectronics is the development of a room temperature single photon detector that simultaneously achieves high count rate, low timing jitter, low dark count rate, and photon-number-resolution. Room temperature single photon detectors have been developed in a range of materials and platforms \cite{Hadfield2009, Zhang2015}, but their performance remains limited by the need to concurrently design optical absorption and electrical readout mechanisms \cite{Ma2015}. Jitter performance in avalanche photodiodes (APDs), for example, is limited by the inhomogeneous travel time of carriers, while thermal noise and electronic defects within the amplification region contribute to high false detection rates, a phenomenon which is particularly significant in non-Si APDs \cite{Hadfield2009, Eisaman2011}. Despite decades of development of passive and active reset mechanisms, reset times are also still long --- typically tens to hundreds of nanoseconds \cite{Hadfield2009,Zhang2015}. Alternatively, superconducting single photon detectors enable high detection efficiency, low dark count rates, and few-ps jitter, but require cryogenic cooling and have limited count rates due to their long dead times \cite{Natarajan2012, Dauler2014, Korzh}.

These examples of state-of-the-art photodetectors illustrate the limitations inherent to amplification in the electronic domain: high thermal noise as well as slow carrier and amplifier response times. Here, we propose a new class of room temperature semiconductor photodetectors that addresses these limitations by realizing single photon amplification in the optical domain. This readout technique retains the benefits of an optical channel: negligible thermal noise, large bandwidth, and low-loss transmission. 

\begin{figure}[t]
\includegraphics[width=0.45\textwidth]{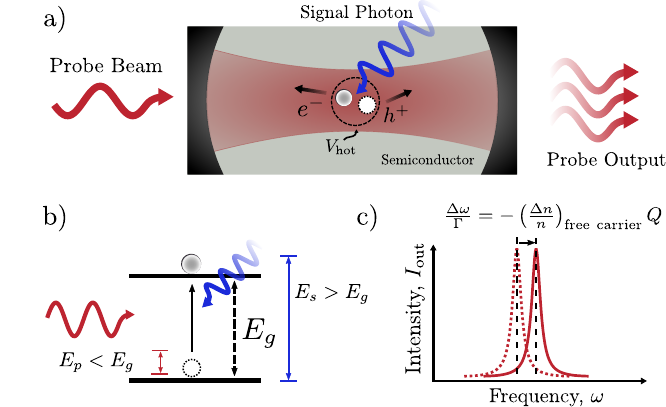}
\caption{Ultrafast all-optical detection of single photons. (a) A signal photon is absorbed in a photonic nanocavity, leading to the generation of a photo-excited charge carrier pair within a ``hot spot" volume $V_\text{hot}$. (b) A sub-bandgap optical probe interrogates the cavity, and is phase shifted as a result of the resonance shift (c) produced by the presence of the additional free carrier.}
\centering
\label{fig: basics}
\end{figure}

Fig.~\ref{fig: basics}(a) illustrates the concept. The absorption of a signal photon in a semiconductor optical cavity excites a free electron-hole charge carrier pair (Fig.~\ref{fig: basics}(b)), which nearly instantaneously shifts the solid-state medium's refractive index --- and in turn the cavity's resonant frequency --- through the free carrier dispersion effect (Fig. 1(c)). The resulting change in resonant frequency imparts a phase shift on a transmitted probe field that can be measured using heterodyne or homodyne detection with common high-speed p-i-n photodetectors \cite{Piels2016, DeRose2011, Chen2017}. Because a single photogenerated electron-hole pair can scatter multiple probe photons, this process produces all-optical gain. The change in cavity transmission induced by a single signal photon can therefore be converted into a strong probe signal at the homodyne receiver. Here, we analyze this all-optical amplification process and determine the experimental requirements for efficient single photon detection.

\section{Free carrier dispersion due to a single electron-hole pair}
\vspace{-10pt}
Suppose that a single photon is absorbed at the center of a cavity as shown in Fig. \ref{fig: basics}(a). According to first order perturbation theory, the photo-excited free electron-hole pair causes a fractional resonance shift \cite{Joannopoulos}
\begin{equation}
\frac{\Delta\omega}{\omega_0}\approx-\frac{1}{2}\frac{\int d^3\vec{r}~\Delta\epsilon(\vec{r})|\vec{E}(\vec{r})|^2}{\int d^3\vec{r}~\epsilon(\vec{r})|\vec{E}(\vec{r})|^2}
\label{eq:perturb}
\end{equation}
due to the permittivity shift $\Delta\epsilon(\vec{r})$ within the electric field profile $\vec{E}(\vec{r})$. In accordance with the Drude model, we assume that the fractional index change $\Delta n/n \approx \Delta\epsilon/2\epsilon$ is directly proportional to the free carrier density. Therefore, if the free carrier pair is confined within a small ``hot spot" volume $V_\text{hot}$ (over which $\vec{E}$ can be assumed to be constant) near the cavity's peak energy, Eqn.~\ref{eq:perturb} simplifies to
\begin{equation}
\frac{\Delta\omega}{\omega_0}\approx\frac{\gamma}{V_\text{eff}},
\end{equation}
where we have introduced the standard optical mode volume $V_\text{eff} =  \int d^3\vec{r}~\epsilon(\vec{r})|\vec{E}(\vec{r})|^2/\text{max}\left\lbrace\epsilon|E|^2 \right\rbrace$  \cite{Note1} and the ``effective scattering volume" $\gamma$ as the constant of proportionality between $|\Delta n/n|$ and carrier density ($1/V_\text{hot}$). This result is identical to the frequency shift generated from a uniform carrier density $1/V_\text{eff}$ throughout the mode volume. The resulting fractional resonance shift with respect to the linewidth $\Gamma$ for a cavity with quality factor $Q=\omega_0/\Gamma$ is
\begin{equation}
\frac{\Delta\omega}{\Gamma}\approx \gamma\frac{Q}{V_\text{eff}}.
\end{equation}
In other words, for any given $\gamma$, a high $Q/V_\text{eff}$ ratio is desired. Silicon photonic crystal (PhC) cavities are therefore an ideal candidate, as recent fabrication advances have enabled cavities with intrinsic $Q$s of $\resim 10^7$ with $V_\text{eff} \sim (\lambda/n)^3$ \cite{Sekoguchi2014, Asano2017a} and alternatively $Q$s of $\resim 10^5$ with mode volumes reaching $\resim 10^{-3}~(\lambda/n)^3$ \cite{Hu2018}. 

In silicon, $\gamma$ can be approximated in two ways. The Drude model in the high frequency limit yields $\gamma = (q_e^2/2n_\text{Si}^2\epsilon_0\omega^2)\left[1/m_e^*+1/m_h^* \right]$ \footnote{See attached supplemental material for further discussion.}, where $q_e$ is the electron charge, $\epsilon_0$ is the vacuum permittivity, $\omega$ is the probe frequency, and $m_e^*$ and $m_h^*$  are the effective masses of the electron and hole, respectively. Given the effective masses $m_e^* = 0.26 m_e$, $m_h^* = 0.39 m_e$ of free carriers in undoped silicon at room temperature \cite{Jacoboni1977, Soref1987}, we find $\gamma \approx 4.3\times10^{-9}~ (\lambda/n)^3$ if a $2.3~\upmu$m probe wavelength is used to eliminate two photon absorption. Alternatively, $\gamma$ can be approximated from the empirical formula \cite{Soref1987, Nedeljkovic2011}
\begin{equation}
\Delta n_\text{Si} = -p(\lambda)[n_e\cdot\text{ cm}^3]^{q(\lambda)}-r(\lambda)[n_h\cdot\text{ cm}^3]^{s(\lambda)},
\label{eqn:soref}
\end{equation}
where $n_e=1/V_\text{hot}$ ($n_h$) is the free electron (hole) density, and $p$, $q$, $r$, and $s$, are the wavelength ($\lambda$) dependent coefficients tabulated in \cite{Nedeljkovic2011}. Eqn.~\ref{eqn:soref} follows from absorption measurements in silicon for wavelengths between 1.3 and 14 $\upmu$m. Linearizing this model about $V_\text{eff}$, we find $\gamma\approx 1.1\times 10^{-8}~ (\lambda/n)^3$ , a factor of $\resim$3 different from the previous estimate.

Both values indicate that a linewidth-order frequency shift requires a quality factor on the order of $Q=V_\text{eff}/\gamma\sim 10^7$ for a probe optical mode volume $V_\text{eff} \sim 10^{-1}~(\lambda_\text{0}/n_\text{Si})^3$, or alternatively $Q\sim 10^5$ for $V_\text{eff} \sim 10^{-3}~(\lambda_\text{0}/n_\text{Si})^3$. As discussed below, optimization of the cavity architecture enables high signal-to-noise ratio (SNR) homodyne detection of the output probe signal with a fractional linewidth shift, even further reducing the necessary $Q$. The approach is therefore applicable to both standard diffraction-limited PhC cavities, as well as state-of-the-art subwavelength-confined nanocavities \cite{Hu2018}. Our subsequent analysis assumes an index change provided by Eqn.~\ref{eqn:soref} due to the experimentally observed nonlinear scaling with respect to carrier density.

Any index change induced by free carrier dispersion is accompanied by a corresponding loss: free carrier absorption (FCA). The associated absorptive quality factor $Q_\text{abs}\approx \lambda\Delta\alpha/2\pi n$ \cite{Xu2009} for an additional FCA loss of $\Delta\alpha$, is therefore of interest. For a single electron-hole pair confined to within the suggested modes volume in silicon, $Q_\text{abs} > 10^6$ \cite{Nedeljkovic2011, Note1}. Since we consider cavities with intrinsic quality factors on the order of $10^5$, we ignore this effect.

\begin{figure*}
\includegraphics[width=\textwidth]{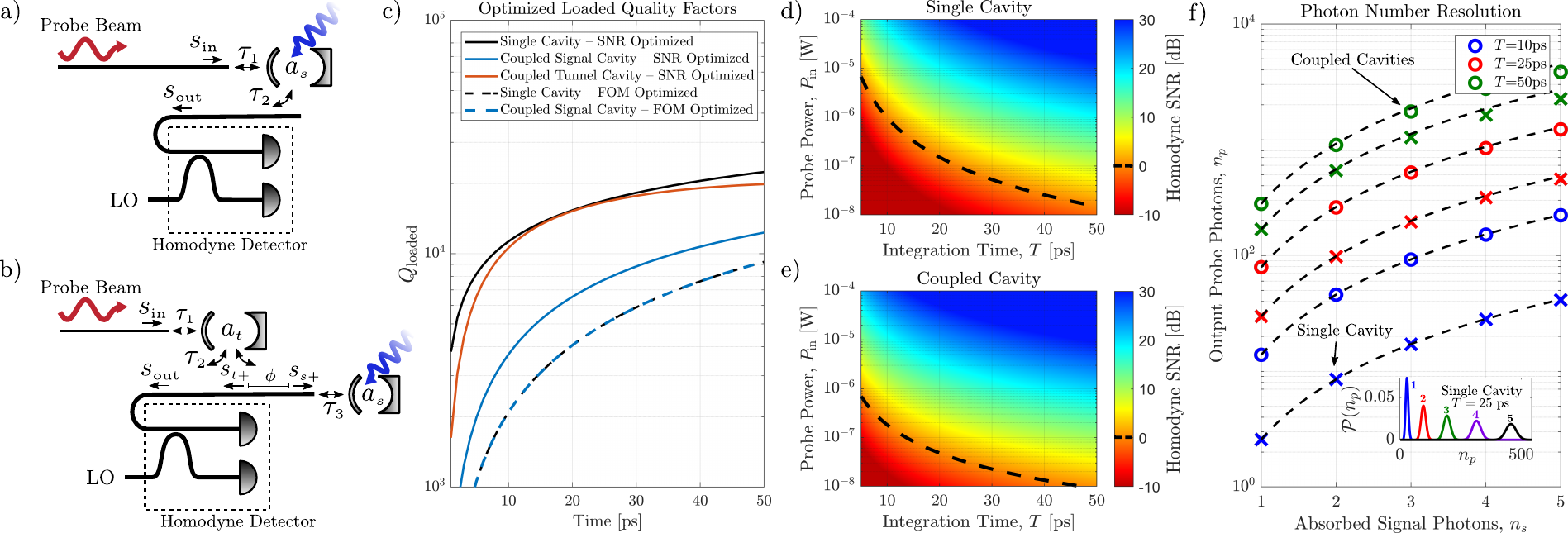}
\caption{Single (a) and coupled cavity (b) architectures for all-optical single photon detection. The signal-to-noise ratio (SNR, Eqn.~\ref{eq:snr}) and figure-of-merit (FOM, Eqn.~\ref{eqn:fom}) are optimized with respect to the coupling rates depicted in (a) and (b) (assuming an intrinsic cavity quality factor of $10^5$, a 2.3 $\upmu$m probe wavelength, and a cavity mode volume of $V_\text{eff}=10^{-3}~(\lambda/n)^3$), yielding the optimal loaded cavity quality factors illustrated in (c). The resulting optimized SNR of a homodyne measurement on the displaced probe signal is shown in (d) and (e) for the single and coupled cavity configurations, respectively. Since the SNR is proportional to the square of the frequency shift induced through free carrier dispersion, photon number resolution --- as illustrated by the scaling characteristics in (f) for a 10 $\upmu$W probe power --- is achievable. Black dashes show that the trend is well fitted by considering the biexponential behavior of Eqn.~\ref{eqn:soref}. The inset depicts the well-separated Poissonian distributions of output probe photon number $n_p$ generated from the absorption of $n_s$ signal photons.}
\centering
\label{fig: results}
\end{figure*}

Other nonlinearities, such as the optical Kerr effect, could be similarly enhanced within these PhC cavities, as the intra-cavity intensity scales with $Q/V_\text{eff}$ \cite{Soljacic2004a, Nozaki2010a}. However, free carrier nonlinearities based on real transitions, while incoherent, are significantly stronger than those resulting from virtual transitions. The effect is also broadband in semiconductors, as any absorbed signal photon produces a free electron-hole pair. Free carrier nonlinearities have previously enabled all-optical switching at GHz speeds with sub-femtojoule ($\resim10^4$ photons) switching energies \cite{Nozaki2010a}. While switching requires $\Delta\omega\sim\Gamma$, we show that photodetection can be achieved with $\Delta\omega\ll\Gamma$, which reduces the required input energy to the single photon level for recently developed high-$Q/V_\text{eff}$ PhC cavities.

\section{Detector Optimization}
\vspace{-10pt}
We analyze the two photodetection architectures shown in Fig.~\ref{fig: results}(a,b) using a temporal coupled mode theory approach \cite{Haus1984, Fan2003, Suh2004}. The simplest configuration (Fig.~\ref{fig: results}(a)) consists of: 1) a single signal cavity, similar to that of Fig.~\ref{fig: basics}, to amplify the phase shift generated by the photoexcited charge carriers; and 2) a homodyne receiver to measure the phase shift of the probe field leaving the cavity. The temporal evolution of the cavity field amplitude $a_s$, assuming input and output probe coupling rates $1/\tau_1$ and $1/\tau_2$ respectively, is governed by the characteristic equation \cite{Haus1984}
\begin{equation}
\frac{da_s}{dt}=\left(j\omega_0-\frac{1}{\tau_s}\right)a_s+j\sqrt{\frac{2}{\tau_1}}s_\text{in},
\end{equation}
where $|s_\text{in}|^2$ is the input power, $|a_s|^2=a_s^*a_s$ is the intra-cavity energy, $\omega_0=\omega$ is the cavity resonant frequency (aligned to the probe frequency $\omega$), and $\tau_s$ is the loaded cavity amplitude decay time. Following the resonance shift $\omega_0\rightarrow\omega_0+\Delta\omega$ generated by the absorption of a signal photon, the shot noise-limited SNR obtained from a homodyne measurement of the displaced output field during a detection window of duration $T$ can be approximated analytically using first order perturbation theory as \cite{Note1}
\begin{equation}
\text{SNR} \approx \frac{8\tau_s^4\Delta\omega^2\left[ 4\tau_s e^{-\frac{T}{\tau_s}}-\tau_s e^{-\frac{2T}{\tau_s}}+2T-3\tau_s\right]}{\hbar\omega_0\tau_1\tau_2}|s_\text{in}|^2,
\label{eq:snr}
\end{equation}
which is directly proportional to the number of probe photons $n_p$ output in response to the resonance shift (SNR=$4n_p$) \cite{Loudon2000}. Eqn.~\ref{eq:snr} illustrates the symmetric dependence of the detection performance upon the input and output probe coupling rates to the signal cavity, $1/\tau_1$ and $1/\tau_2$, respectively. In general, faster coupling rates limit the detector sensitivity but enable the displaced probe signal to be rapidly extracted, while the opposite is true for slow coupling rates. Optimizing the SNR with respect to $\tau_1$ and $\tau_2$ yields the optimum loaded quality factor, $Q_\text{loaded}=\omega_0\tau_s/2$, shown in Fig.~\ref{fig: results}(c) (assuming an intrinsic quality factor $Q_i = 10^5$), which produces the SNR shown in Fig.~\ref{fig: results}(d) for a 2.3 $\upmu$m probe wavelength. 

The results demonstrate that a SNR$\gg$1 is achievable within 50 ps for a sub-$\upmu$W probe power due to the all-optical gain afforded by the detection cavity. Notably, the optimal loaded quality factors ($\resim 10^4$) for these short (sub-50 ps) integration times are much less than $Q_i$, indicating that $Q_i$ can be further reduced without substantial degradation of the detection SNR. Similar results are obtained for $V_\text{eff}=10^{-1} (\lambda/n)^3$ \cite{Note1}; however, the input power required for high-SNR detection within a given time increases.

The coupled-cavity architecture shown in Fig.~\ref{fig: results}(b) can be used for ``cavity dumping" to reduce the required probe power. The evolution of the two cavities, assumed to be resonant with the input probe frequency $\omega$, is described by the coupled differential equations
\begin{align}
\frac{da_t}{dt} &= \left(j\omega-\frac{1}{\tau_t}\right)a_t + j\sqrt{\frac{2}{\tau_1}}s_\text{in}+j\sqrt{\frac{2}{\tau_2}}s_{t+}\label{coupledEquations1}
\\
\frac{da_s}{dt} &= \left(j\omega-\frac{1}{\tau_s}\right)a_s+j\sqrt{\frac{2}{\tau_3}}s_{s+},
\label{coupledEquations2}
\end{align}
where $a_t$ ($a_s$) is the tunnel (signal) cavity amplitude that decays at rate $1/\tau_t = 1/\tau_i+1/\tau_1+2/\tau_2$ ($1/\tau_s = 1/\tau_i+1/\tau_3$) for the coupling times shown in Fig.~\ref{fig: results}(b). The waves $s_{s+}=j\sqrt{2/\tau_2}e^{j\phi}a_t$ and $s_{t+}=j\sqrt{2/\tau_3}e^{j\phi}a_s+j\sqrt{2/\tau_2}e^{j2\phi}a_t$ couple the tunnel and signal cavities, which are separated by a distance corresponding to an effective phase $\phi$. If $\phi=m\pi$ \cite{Note1} for any integer $m$, the wave output from $a_s$ destructively interferes with $s_{t+}$ at $s_\text{out}$, corresponding to the high-$Q$ regime of the two cavity system. The phase shift produced through the absorption of a signal photon within $a_s$ disturbs this interference condition, causing rapid evacuation of the stored cavity field. This effect, analogous to ``cavity dumping" for pulse generation in laser resonators \cite{Pshenichnikov1994}, has been previously implemented to achieve ultrafast tuning of photonic crystal cavities \cite{Tanaka2007} and integrated ring resonators \cite{Xu2007a}. 

Given these considerations, we numerically optimized the coupling rates of Eqns.~\ref{coupledEquations1} and \ref{coupledEquations2} to maximize the SNR in Eqn.~\ref{eq:snr}, yielding the loaded quality factors shown in Fig.~\ref{fig: results}(c) and associated SNR in Fig.~\ref{fig: results}(d). These simulations indicate that the power reduction is proportional to $\resim Q_i/Q_3$, which can be understood as the amplification of stored energy in the high-$Q$ regime.

Furthermore, in the perturbative limit, the functional form of the SNR provided in Eqn.~\ref{eq:snr} shows that $n_p$ is proportional to the square of the resonance shift, and is thereby a function of the number of absorbed signal photons $n_s$. Fig.~\ref{fig: results}(f) demonstrates this scaling for a 10 $\upmu$W input probe power. For small $n_s$, the trend is well fitted by considering the biexponential behavior of the refractive index shift in Eqn.~\ref{eqn:soref}. The growth of $n_p$ with respect to $n_s$ is large enough to enable photon number resolution of the signal field. This is illustrated by the well-separated Poissonian distributions of $n_p$ in the inset of Fig.~\ref{fig: results}(f) for $n_s\in[1,5]$.

Overall, these optimized results demonstrate that a single silicon nanocavity can enable ultrafast, high-efficiency, and even number resolving single photon detection. Cavity dumping in a two-cavity system can reduce the probe field power by over an order of magnitude. This improvement also opens the possibility of monitoring a large array of detectors with a significantly reduced input probe power. Moreover, jitter contributions for the all-optical photodetector are limited to fluctuations in the signal photon absorption time, probabalistic variation in the output probe field, and the jitter of the homodyne photodetectors. As the first two are negligible for high-efficiency detection \cite{Note1}, the achievable jitter is limited by that of the photodetectors used to measure the classical probe field.

\section{Dark Count Contributions}
\vspace{-10pt}
The aforementioned analysis considered the probe laser shot noise as the sole source of noise. In reality, dark counts --- erroneous detection events which occur in absence of a signal beam --- must be carefully considered. Any fast (on the order of the detector integration time $T$) change in the cavity refractive index above $\Delta n$ contributes to the dark count rate, as slow changes can be high-pass filtered. We consider three principal factors: thermal excitation of free carriers, temperature fluctuations of the semiconductor substrate, and multiphoton absorption. While surface defect states may contribute to the dark count rate, we omit this contribution due to the record-low surface recombination velocity of silicon \cite{Yablonovitch1986}. 

Given an intrinsic carrier concentration of $1.5\times 10^{10}~\text{cm}^{-3}$ in pure silicon at 300 K, the mean population within the proposed mode volume $V_\text{eff}=10^{-3}~(\lambda/n_\text{Si})^3$ at a 2.3 $\upmu$m probe wavelength is $\resim4\times 10^{-6}$. The resulting probability of a non-zero thermal carrier population within the optical mode is approximately equal to this mean occupancy \cite{Note1} and corresponds to a 10 kHz dark count rate for a $\resim 40$ GHz detector gating frequency. Cooling to 200 K reduces this dark count rate to a negligible sub-Hz rate \cite{Note1}. The dark count rate from temperature \textit{variation} of the substrate is likewise negligible if the temperature is stabilized to $\Delta T<\Delta n_\text{Si}/\alpha_\text{TO}\sim0.1$ K (where $\alpha_\text{TO}\sim10^{-4}$ K$^{-1}$ is silicon's thermo-optic coefficient) such that thermo-optic refractive index changes are smaller than those induced by a single absorbed photon. Fundamental statistical temperature fluctuations within the cavity --- which typically limit PhC cavity sensitivity \cite{Saurav} --- are an order of magnitude smaller than this value \cite{Note1}, and the stability requirement can therefore be readily achieved with modern PID temperature controllers \cite{Lee2016}.
 
While both thermally induced free carriers and direct index variations due to the thermo-optic effect in silicon can be mitigated with proper environmental control \cite{Note1}, multiphoton absorption (MPA) events --- given the indistinguishably between probe- and signal-induced free carriers within the cavity --- produce a dark count rate that can only be lowered by minimizing the intensity of the probe laser within the signal cavity, and thus inherently reducing the sensitivity of the detector. The overall dark count rate from MPA is
\begin{equation}
R_\text{dark} = \int \frac{\beta_k}{k\hbar\omega}I(\vec{r})^k~ d^3\vec{r},
\end{equation}
where $\beta_k$ is the MPA coefficient for $k$-photon absorption ($k$PA) and $I(\vec{r})$ is the probe intensity at a position $\vec{r}$. Re-expressing this definition in terms of the peak cavity intensity $I_\text{max}=|a_s|^2c/2n_\text{Si}V_\text{eff}$, we find \cite{Note1}
\begin{equation}
R_\text{dark} = \frac{\beta_k}{k\hbar\omega}I_\text{max}^kV_\text{k\text{PA}},
\label{eqn:dcr}
\end{equation}
where the multiphoton absorption mode volume $V_{k\text{PA}}$ is defined as
\begin{equation}
V_{k\text{PA}}\equiv \frac{\int_\text{Si}  \epsilon_\text{Si}^k|\vec{E}(\vec{r})|^{2k}d^3\vec{r}}{\text{max}\left\lbrace \epsilon_\text{Si}^k|\vec{E}|^{2k} \right\rbrace}.
\end{equation}

For the deep subwavelength ($V_\text{eff}\ll (\lambda/n)^3$) ``tip"-based cavities of interest to this study \cite{Hu2016, Choi, Hu2018}, a defect in the cavity geometry yields a localized, high intensity region within the broader diffraction-limited mode size \cite{Boroditsky1998}. We studied the mode profile of the silicon tip cavity in \cite{Choi} to determine the scaling of $V_{k\text{PA}}$ in this case, yielding the mode volumes summarized in Table~\ref{table} \cite{Note1}. A significant volume reduction is achieved for $k>1$, revealing an advantage of the tip-based cavity for low noise, all-optical photodetection --- suppression of higher order noise processes.
\begin{table}
\caption{\label{table} Overview of wavelengths of interest and their associated multiphoton absorption (MPA) parameters for the dominant $k$th order process in the ``tip" defect cavity in \cite{Choi}.}
\begin{ruledtabular}
\begin{tabular}{C{2cm}C{4cm}C{2.3cm}}
Probe Wav. $\lambda$ [$\upmu$m]  & MPA Coefficient (Process) \cite{Bristow2007,Pearl2008a,Gai2013} & $V_{k\text{PA}}/V_\text{eff}$ \\
\hline  \addlinespace[1ex]
2.3 & 2.5$\times 10^{-2}$ cm$^3$/GW$^2$ (3PA) & 7.82$\times 10^{-3}$ \\
3.4 & 2.5$\times 10^{-4}$ cm$^5$/GW$^3$ (4PA) &  2.00$\times 10^{-3}$ \\
4.5 & 1.4$\times 10^{-6}$ cm$^7$/GW$^4$ (5PA) &  6.14$\times 10^{-4}$ \\
\end{tabular}
\end{ruledtabular}
\end{table}
Given the ability to accurately estimate the dark count rate in Eqn.~\ref{eqn:dcr} using the parameters in Table~\ref{table}, we re-optimize the cavity coupling rates with respect to the figure-of-merit 
\begin{equation}
\text{FOM}=\frac{n_p}{|a_s|^2}=\frac{\text{SNR}}{4|a_s|^2},
\label{eqn:fom}
\end{equation}
such that the maximum output is achieved for a given intra-cavity intensity or dark count rate. Since the FOM is independent of input power, both the single and coupled cavity architectures achieve the same optimum value of \cite{Note1}
\begin{equation}
\text{FOM}_\text{opt} \approx 0.381~\Delta\omega^2T^2/\hbar\omega
\end{equation}
for the loaded quality factors shown in Fig.~\ref{fig: results}(c). Eqn.~\ref{eqn:dcr} can then be re-parameterized in terms of this FOM and the shot noise limited detection efficiency $\eta_\text{SN} = 1-\exp(-n_p)$, yielding \cite{Note1}
\begin{equation}
R_\text{dark} \approx \frac{\beta_k}{k\hbar\omega\text{FOM}_\text{opt}^k}\left(\frac{c}{2n_\text{Si}}\right)^k\frac{V_{k\text{PA}}}{V_\text{eff}^k}\ln\left(\frac{1}{1-\eta_{SN}}\right)^k.
\end{equation}
Assuming a linear scaling of $\Delta\omega$ with carrier density as in the Drude model, the dark count rate scales as $\resim V_\text{eff}^{1+k}$, revealing the performance enhancement that can be achieved by minimizing the optical mode volume. Longer wavelengths also reduce multi-photon absorption.
\begin{figure}
\includegraphics[width=.5\textwidth]{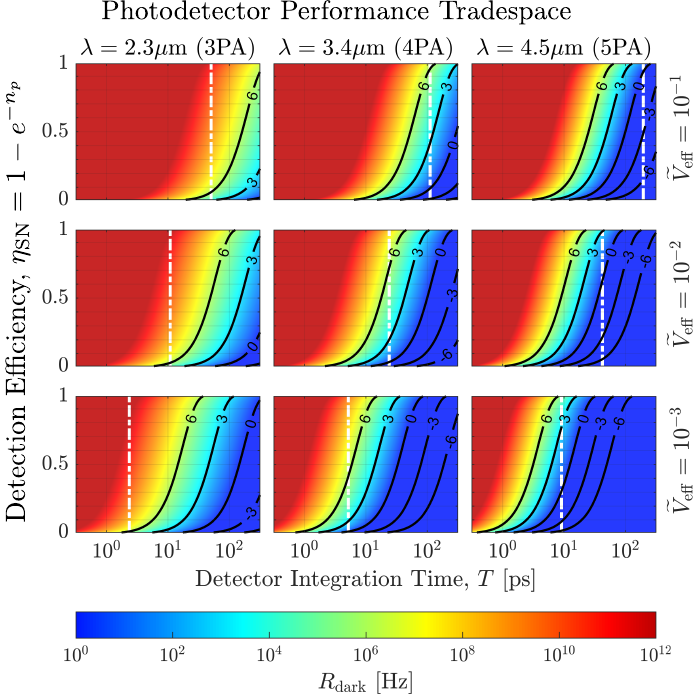}
\caption{Optimized multiphoton absorption-induced dark count rates $R_\text{dark}$ as a function of detection efficiency $\eta_\text{SN}$ and integration time $T$ for the tip defect cavity parameters in Table~\ref{table}. Rows indicate different optical mode volumes $\widetilde{V}_\text{eff} = V_\text{eff}/(\lambda/n_\text{Si})^3$ (from varying fabrication precision \cite{Choi}, for example), while columns correspond to probe wavelengths associated with different orders of multiphoton absorption (MPA). The vertical white lines indicate the expected lifetime of free carriers within the cavity mode volume as dictated by ambipolar diffusion \cite{Nozaki2010a}.}
\centering
\label{fig: mpatradespace}
\end{figure}
The optimized detection tradespaces --- dark count rate as a function of the desired detection efficiency $\eta_\text{SN}$ and integration time $T$ --- for the cavity parameters in Table~\ref{table} are plotted in Figure~\ref{fig: mpatradespace} for various effective volumes $\widetilde{V}_\text{eff} = V_\text{eff}/(\lambda/n_\text{Si})^3$. The results confirm the aforementioned suppression of dark counts at small mode volumes and long wavelengths. For example, the baseline parameters of the previous section ($\lambda=$2.3$\upmu$m and $\widetilde{V}_\text{eff}=10^{-3}$) result in a dark count rate on the order of 100 kHz given a 50\% detection efficiency and 20 ps integration time; however, this rate can be reduced to below 0.01 Hz at $\lambda=$4.5 $\upmu$m.

While optical switching experiments \cite{Tanabe2008, Nozaki2010a, TurnerFoster2010} seek to minimize the ps-scale diffusion times of photogenerated carriers to maximize the achievable switching frequency, Fig.~\ref{fig: mpatradespace} demonstrates the advantage of reduced dark counts with an extended detection time. The maximum integration time is limited by carrier diffusion for small mode volume PhC cavities \cite{Tanabe2008, Aldaya2017}. To extend this time (estimated by the vertical dashed lines in Fig.~\ref{fig: mpatradespace} assuming an ambipolar diffusion constant $D_\text{am}=19$ cm$^2$/s in silicon \cite{Nozaki2010a}), charge confinement techniques, such as the incorporation of a double heterostructure \cite{Alferov2001}, may be necessary. Even in the absence of charge confinement, the performance tradespace illustrates that an appropriate combination of probe wavelength and cavity volume can be selected to achieve the desired detection characteristics. Most importantly, these optimized metrics demonstrate the ability to realize efficient single photon detection within tens of picoseconds using experimentally achievable parameters.

The resulting overall detection efficiency $\eta=\eta_\text{abs}\eta_\text{SN}$ is limited by the absorption efficiency $\eta_\text{abs}$ of the incident signal photon within the mode volume of the probe cavity. For standard suspended silicon photonic crystal cavities with $\tilde{V}_\text{eff}\sim 1$ at $\lambda = 4.5~\upmu$m, numerical simulations yield a peak efficiency $\eta_\text{abs}\sim 0.6$ for focused visible light, and $\eta_\text{abs} \sim 1$ could be achieved by incorporating anti-reflection and reflection coatings on the top and bottom surfaces of the cavity, respectively \cite{Note1}. However, absorption within the subwavelength-confined mode volumes ($\tilde{V}_\text{eff}=10^{-3}$) characteristic of tip defect cavities is limited to $\eta_\text{abs}\sim 0.15$ \cite{Note1}. Future work will therefore consider techniques for localized signal light absorption within ultrasmall mode volume nanocavities. Possible approaches include the design of a doubly-resonant cavity for the probe and signal fields \cite{Rivoire2011, Hueting2014} or the incorporation of a selective absorber, such as a buried heterostructure \cite{Matsuo2010a}, at the center of the cavity. 

\section{Conclusion} In summary, we have analyzed a new concept for single photon detection based on all-optical gain in a nanocavity system. The proposed amplification mechanism can be of use in a range of all-optical signal processing applications, and in particular opens the possibility of room-temperature single photon detection. A single cavity suffices to implement the scheme, and interference with a second cavity can reduce the required input power by orders of magnitude. Multiphoton absorption is a dominant noise process, but the resulting dark count rate can be minimized through a combination of a long-wavelength probe field, a subwavelength confining nanocavity, squeezed light \cite{Note1}, or the use of a wide bandgap probe cavity material. The proposed dielectric cavity-based system for room temperature, low power, ultrafast single photon detection would prove useful in a wide range of photonic technologies.

The authors thank H. Choi and C. Peng for providing the mode profiles evaluated in this study. C.P. was supported by the Hertz Foundation Elizabeth and Stephen Fantone Family Fellowship and the MIT Jacobs Presidential Fellowship. M.H. was supported by Villum Fonden. R.H. is supported by an appointment to the IC Postdoctoral Research Fellowship Program at MIT, administered by ORISE through U.S. DOE/ODNI. This work was supported in by by the AFOSR MURI for Optimal Measurements for Scalable Quantum Technologies (FA9550-14-1-0052) and by the AFOSR program FA9550-16-1-0391, supervised by Gernot Pomrenke.

\vspace{-16pt}

%

\end{document}


\widetext
\clearpage
\begin{center}
\textbf{\large Supplemental Materials: Single-Photon Detection by Cavity-Assisted All-Optical Gain}
\end{center}
\setcounter{equation}{0}
\setcounter{figure}{0}
\setcounter{table}{0}
\setcounter{page}{1}
\makeatletter
\renewcommand{\theequation}{S\arabic{equation}}
\renewcommand{\thefigure}{S\arabic{figure}}
\renewcommand{\citenumfont}[1]{S#1}
\renewcommand{\bibnumfmt}[1]{[S#1]}

\section{I. Free Carrier Nonlinearities}

\begin{figure*}[h!]
\includegraphics[width=.4\textwidth]{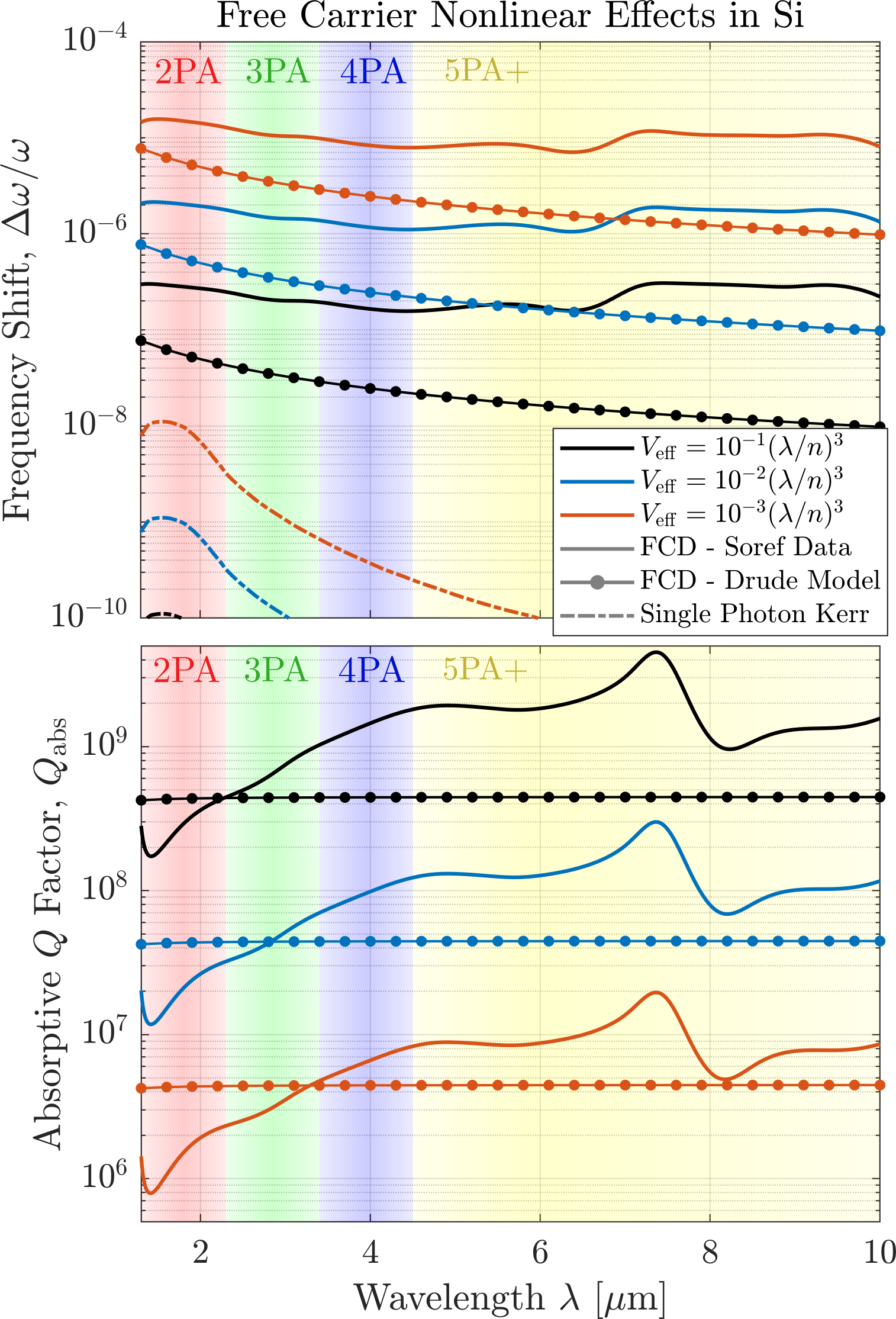}
\caption{Dispersive and absorptive characteristics of the free carrier nonlinearity in the limit of a single photoexcited charge pair. The perturbation of the real component of the refractive index leads to a fractional frequency shift $\Delta\omega/\omega\approx -\Delta n_\text{Si}/n_\text{Si}$, while free carrier absorption induces an additional loss $\Delta\alpha$, corresponding to an absorptive quality factor $Q_\text{abs}\approx \Delta\alpha\lambda/2\pi n_\text{Si}.$ Both quantities can be estimated with the Drude model (dotted lines) or using the empirical models provided in \cite{Nedeljkovic2011}. Regardless of the approximation method, the strength is found to be much larger than that of a single photon Kerr nonlinearity (dashed).}
\centering
\label{figS1:nonlinearities}
\end{figure*}

Figure \ref{figS1:nonlinearities} provides an overview of the absorptive and dispersive effects of a single photon-induced free carrier nonlinearity as obtained from: 1) a Drude model estimate, and 2) extrapolation from the experimentally derived models in \cite{Soref1987, Nedeljkovic2011}. 

Free carrier dispersion yields a complex permitivity shift $\Delta \epsilon$ that, according to first order perturbation theory \cite{Joannopoulos}, induces a fractional resonance shift
\begin{equation}
\label{eq:dom}
\frac{\Delta \omega}{\omega_0} = -\frac{1}{2}\frac{\int_{-\infty}^{\infty} d^3\vec{r}~\Delta\epsilon(\vec{r})|\vec{E}(\vec{r})|^2}{\int_{-\infty}^{\infty} d^3\vec{r}~\epsilon(\vec{r})|\vec{E}(\vec{r})|^2}.
\end{equation}
Assuming a uniform, perturbative index change within a small ``hot spot" volume $V_\text{hot}$ near the cavity's peak energy density at $\vec{r}=\vec{r}_\text{max}$, this simplifies to
\begin{align}
\frac{\Delta\omega}{\omega_0}&\approx-\frac{\int_{V_\text{hot}} d^3\vec{r}~\frac{\Delta n(\vec{r})}{n(\vec{r}_\text{max})}\big|\frac{\vec{E}(\vec{r})}{\vec{E}(\vec{r}_\text{max})}\big|^2}{\frac{\int d^3\vec{r}~\epsilon(\vec{r})|\vec{E}(\vec{r})|^2}{\text{max}\left\lbrace\epsilon|E|^2 \right\rbrace}} \approx -\frac{\Delta n}{n}\frac{V_\text{hot}}{V_\text{eff}}=\frac{\gamma}{V_\text{hot}}\frac{V_\text{hot}}{V_\text{eff}}=\frac{\gamma}{V_\text{eff}},
\end{align}
where we have approximated $\Delta n \approx (\Delta\epsilon/2\epsilon)n$, defined $-\gamma$ as the constant of proportionality between $\Delta n/n$ and carrier density ($1/V_\text{hot}$), and introduced the standard effective optical mode volume
\begin{equation}
V_\text{eff} =  \frac{\int d^3\vec{r}~\epsilon(\vec{r})|\vec{E}(\vec{r})|^2}{\text{max}\left\lbrace\epsilon|E|^2 \right\rbrace}.
\end{equation}
Note that the validity of this analysis requires the absorption of the signal photon and the resulting index change to occur near the cavity's peak energy density (i.e. well within the optical mode volume $V_\text{eff}$).

In silicon, previous analyses have demonstrated that $\Delta n$ is primarily a result of coulomb interactions with the free carrier, while Burstein-Moss bandfilling is negligible \cite{Soref1987, Nedeljkovic2011}.  Therefore, a simple Drude model analysis of this process estimates that the injection of a single free carrier pair into a volume $V_\text{hot}$ will shift the complex permittivity $\epsilon$ by
\begin{equation}
\label{deMain}
\Delta \epsilon = \frac{q_e^2}{j\omega\epsilon_0V_\text{hot}}\left[ \frac{\tau_e}{m^*_e(1+j\omega\tau_e)} + \frac{\tau_h}{m^*_h(1+j\omega\tau_h)} \right]
\end{equation}
where $q_e$ is the electron charge, $\epsilon_0$ is the vacuum permittivity, $\omega$ is the probe frequency, and $\tau$ and $m^*$  are the effective mass and mean collision time of the free charge carriers (electrons for ``$e$" subscripts, and holes for ``$h$" subscripts), respectively.  The mean collision time governs the resulting behavior, and can be approximated using the experimentally measured mobilities $\mu_e \approx 1.5 \times 10^3 \text{ cm}^2 \text{ V}^{-1}\text{ s}^{-1}$, $\mu_h \approx 500 \text{ cm}^2 \text{ V}^{-1}\text{ s}^{-1}$ and effective masses $m_e^* = 0.26 m_e$, $m_h^* = 0.39 m_e$ of free carriers in undoped silicon at room temperature,  which yields $\tau_e = \mu_em_e^*/q_e \approx 0.22 \text{ ps}$ and $\tau_h = \mu_hm_h^*/q_e \approx 0.11 \text{ ps}$ \cite{Jacoboni1977, Soref1987}. Assuming a probe wavelength of $\lambda_0=2.3\mu$m is used to avoid two photon absorption, the high-frequency limit $\omega\tau \gg 1$ of Equation \ref{deMain} for a weakly absorbing medium yields a frequency shift governed by
\begin{equation}
\gamma = -\frac{1}{V_\text{hot}}\frac{\Delta n_\text{Si}}{n_\text{Si}} = \frac{q_e^2}{2n_\text{Si}^2\epsilon_0\omega^2}\left[\frac{1}{m_e^*}+\frac{1}{m_h^*} \right]
\end{equation}
and an additional free carrier absorption loss
\begin{equation}
\Delta \alpha = \frac{2\pi}{\lambda_0} \frac{\epsilon_0 n_\text{Si} V_\text{hot}}{q_e^2}\left[\tau_em_e^*+\tau_hm_h^* \right]\omega^3.
\end{equation}
For simplicity, we assume $V_\text{hot} = V_\text{eff}$ for loss calculations. The results demonstrate that confining the free charge carrier pair within an experimentally-feasible probe optical mode volume $V_\text{eff} = 10^{-3}(\lambda_\text{0}/n_\text{Si})^3=2.8\times10^{-4}\mu\text{m}^3$ lends an overall index change of $\Delta n_\text{Si} \approx -1.5\times10^{-5}$. 

At a given resonant wavelength $\lambda$, the corresponding loss $\Delta \alpha$ from free carrier absorption can be represented with an additional absorptive quality factor \cite{Xu2009}
\begin{equation}
Q_\text{abs}\approx \frac{\Delta\alpha\lambda}{2\pi n_\text{Si}}.
\end{equation}
$Q_\text{abs}$ can be estimated directly from experimental absorption data, or from the Drude model, where $Q_\text{abs}=\omega^3\tau n_\text{Si}/\omega_p^2$ for a cavity resonance at $\omega$ in a material with a plasma frequency $\omega_p=\sqrt{nq_e^2/m\epsilon_0}$ -- resulting from the presence of a carrier density $n$ of individual charge $q_e$ and mass $m$ -- and mean collision time $\tau$ as determined by the carrier mobility. The results of Figure \ref{figS1:nonlinearities} illustrate that this additional loss is negligible for all cavity mode volumes of interest, as $Q_\text{abs}$ is much larger than the desired intrinsic cavity quality factors $Q_i$ ($\resim 10^5$).

The free carrier nonlinearity can also be compared to a single photon Kerr nonlinearity, which leads to a fractional resonance shift \cite{Choi}
\begin{equation}
\left(\frac{\Delta\omega}{\omega_0}\right)_\text{Kerr}=-\frac{3\chi^{(3)}(\omega_0)}{4\epsilon_0n_\text{Si}^4}\left(\frac{U_\text{sig}}{V_\text{eff}} \right)\frac{V_\text{2PA}}{V_\text{eff}}
\end{equation}
for a third order nonlinear susceptibility $\chi^{(3)}$ and a signal photon  energy $U_\text{sig}$. The mode volume $V_\text{2PA}$ is defined in Section IV, and the cavity profile shown in Fig.\ref{figS:volScaling} yields $V_\text{2PA}/V_\text{eff}\approx 0.04$. We estimate $\chi^{(3)}(\omega)$ using the mean fit provided in \cite{Hon2011}, which results in the values plotted in Figure \ref{figS1:nonlinearities} for a band-edge ($U_\text{sig}=$1.1 eV) signal photon in silicon. The results illustrate that, as expected due to the virtual transition states involved with the Kerr effect as opposed to real transitions for free carrier nonlinearities, the effect is over an order of magnitude weaker than free carrier dispersion. The Kerr effect is also weaker at longer wavelengths, thus restricting the feasible range of probe wavelenths. Similar conclusions are provided in \cite{Soref1987, Nozaki2010a}, and we therefore focus on free carrier nonlinearities for the photodetection scheme.

\section{II. Single Cavity Theory}
\label{sec:theory}

\subsection{Architecture and Performance Metrics}

Following the time-dependent formulation of optical cavity fields provided in \cite{Haus1984}, the time evolution of a general cavity with $N$ input-output ports can be described by the differential equation
\begin{equation}
\frac{da}{dt} = j\omega_0a-\left(\frac{1}{\tau_i}+\sum_{n=1}^N \frac{1}{\tau_n} \right)a +\sum_{n=1}^N \kappa_n s_{n+},
\label{de}
\end{equation}
where $a$ describes the field of a cavity with resonant frequency $\omega_0$, $\tau_i$ and $\tau_n$ are the decay times due to intrinsic loss and external coupling at port $n$, respectively, and $\kappa_n^2$ is power coupling coefficient at port $n$. It should be noted that the cavity amplitude $a$ is energy normalized, such that $|a|^2$ is the total energy within the cavity, while the input/output field amplitudes $s_{n\pm}$ are power normalized (power $= |s_{n\pm}|^2$).

\begin{figure}[h!]
\includegraphics[width=.5\textwidth]{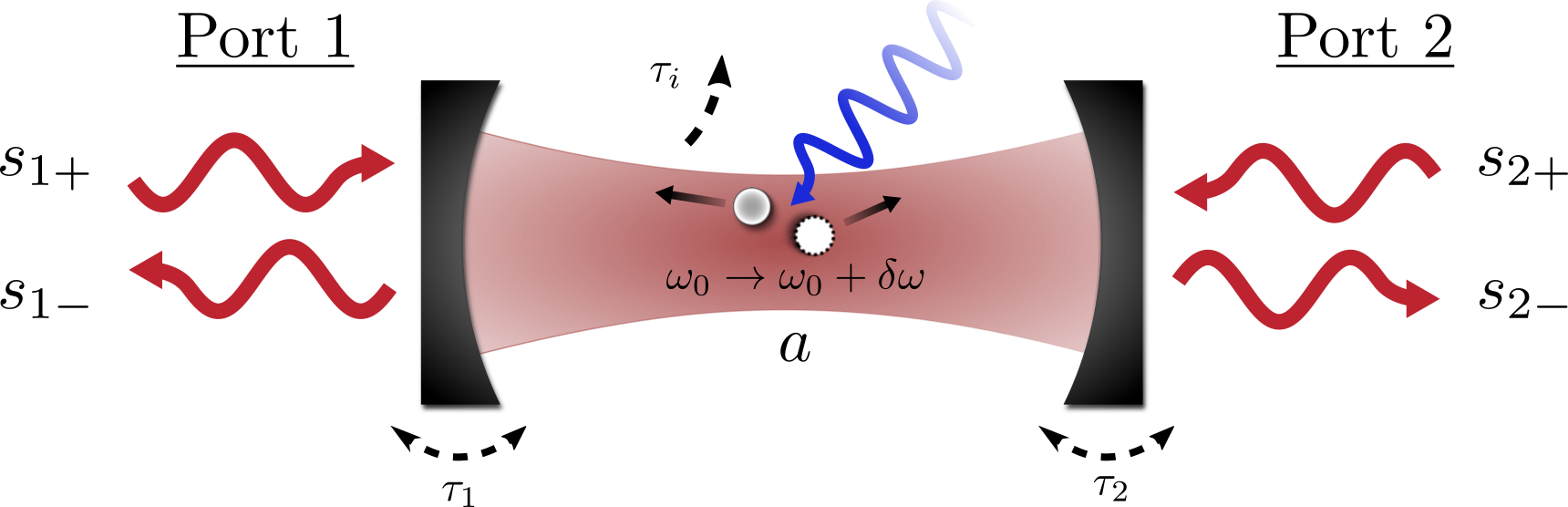}
\caption{Diagram of the single cavity photodetector. An incident signal photon (blue) is absorbed within the cavity, causing a small shift in the cavity resonant frequency $\omega_0$, which in turn modifies the field amplitudes $a$ and $s_{n\pm}$.}
\label{fig: schematic}
\end{figure}

Here, we analytically derive the field evolution of a cavity with two input/output ports in response to the instantaneous change in the resonant frequency produced by the absorption of a single photon in order to determine the system's performance as an all-optical single photon detector. The variables of interest for the two-port cavity system are schematically depicted in Fig. \ref{fig: schematic}.

Assuming a two-port architecture which is only excited at port 1, Equation \ref{de} simplifies to \cite{Haus1984}
\begin{equation}
\frac{da}{dt} = j\omega_0a-\frac{1}{\tau}a + \sqrt{\frac{2}{\tau_1}} s_{1+},
\label{de3}
\end{equation}
where we have defined the composite decay rate $1/\tau = 1/\tau_i+1/\tau_1+1/\tau_2$ and substituted $\kappa_1 = \sqrt{2/\tau_1}$.

Prior to the perturbation of the cavity resonance frequency, the steady state cavity amplitude -- assuming an input excitation $s_{1+}=s_+e^{j\omega t} + \text{c.c.} $ -- is found to be
\begin{equation}
a = \frac{\sqrt{2/\tau_1} s_{1+}}{1/\tau+j(\omega-\omega_0)} = \frac{\sqrt{2/\tau_1} s_+}{1/\tau+j(\omega-\omega_0)}e^{j\omega t} + \text{c.c.}
\end{equation}
with a total cavity energy
\begin{equation}
|a|^2 = a^*a = \frac{2/\tau_1 |s_+|^2}{1/\tau^2+(\omega-\omega_0)^2}
\end{equation}
that is directly proportional to the input power $|s_+|^2$.

This steady state solution can then be used to find the field changes that occur in response to the instantaneous resonant frequency change $\omega_0\rightarrow\omega_0'\equiv\omega_0+\Delta\omega$ by assuming a perturbative solution $a' = a+\delta a$. The differential equation thus becomes
\begin{equation}
\frac{d}{dt}(a+\delta a) = j(\omega_0+\Delta\omega)(a+\delta a)-\frac{1}{\tau}(a+\delta a) + \sqrt{\frac{2}{\tau_1}} s_{1+},
\end{equation}
which, by eliminating the steady state relationship in Equation \ref{de3}, simplifies to 
\begin{equation}
\frac{d}{dt}\delta a = j\omega_0'\delta a -\frac{1}{\tau}\delta a + j\Delta\omega a.
\end{equation}
Solving this result subject to the initial condition $\delta a|_{t=0}=0$ yields
\begin{equation}
\delta a = \frac{\sqrt{2/\tau_1}\tau^2\Delta\omega\left[ e^{(j\omega_0'-1/\tau)t}-e^{j\omega t}\right]}{\left[1+j\tau(\omega-\omega_0)\right]\left[j-\tau(\omega-\omega_0')\right]}s_{1+}.
\end{equation}
The change in cavity energy is therefore
\begin{equation}
|\delta a|^2 = \frac{2\tau^4\Delta\omega^2\left[1+e^{2t/\tau}-2e^{t/\tau}\cos((\omega-\omega_0')t)\right]e^{-2t/\tau}}{\tau_1\left[ (1+\tau^2(\omega_0'-\omega)^2)(1+\tau^2(\omega-\omega_0)^2 )\right]}|s_+|^2.
\end{equation}
which can be simplified to
\begin{equation}
|\delta a|^2 = \frac{2\tau^4\Delta\omega^2\left[1+e^{2t/\tau}-2e^{t/\tau}\cos(\Delta\omega t)\right]e^{-2t/\tau}}{(1+\tau^2\Delta\omega^2 )^2\tau_1}|s_+|^2
\end{equation}
by assuming resonant excitation ($\omega = \omega_0$). The power transmitted from port 2, $|s_{2-}|^2$, is directly proportional to the cavity energy, with $\kappa_2$ serving as the constant of proportionality \cite{Haus1984}. Assuming a first order solution in $\delta a$ and $\delta s_{2-}$, it can be easily shown that this relationship implies
\begin{equation}
|\delta s_{2-}|^2 \approx \frac{2}{\tau_2}|\delta a|^2 = \frac{4\tau^4\Delta\omega^2\left[1+e^{2t/\tau}-2e^{t/\tau}\cos(\Delta\omega t)\right]e^{-2t/\tau}}{(1+\tau^2\Delta\omega^2 )^2\tau_1\tau_2}|s_+|^2.
\label{outputpwr}
\end{equation}

The number of output probe photons produced within an integration time $T$ in response to the absorption of a single signal photon -- which we term the \textit{single photon gain} -- is simply the integral of Eqn. \ref{outputpwr} normalized to the probe photon energy $\hbar\omega_0$, resulting in
\begin{align}
n_p &= \int_0^T |\delta s_{2-}|^2 dt \bigg/ \hbar\omega_0 \\
\begin{split}
n_p	&= 2\tau^4\Delta\omega^2\bigg\lbrace 4\tau e^{T/\tau}\left[ \cos(\Delta\omega T)-\tau\Delta\omega\sin(\Delta\omega T)\right]\\ &~~~ -\tau(1+\tau^2\Delta\omega^2)+e^{2T/\tau}\left[\tau^2\Delta\omega^2(\tau+2T)+2T-3\tau \right] \bigg\rbrace e^{-2T/\tau}|s_+|^2\bigg/ \hbar\omega_0(1+\tau^2\Delta\omega^2)^2\tau_1\tau_2
\end{split}
\end{align}
Given a sufficiently small resonant frequency change such that $\tau\Delta\omega\ll 1$, the number of output probe photons can be approximated as
\begin{equation}
\boxed{n_p \approx \frac{2\tau^4\Delta\omega^2\left[ 4\tau e^{-T/\tau}-\tau e^{-2T/\tau}+2T-3\tau\right]}{\hbar\omega_0\tau_1\tau_2}|s_+|^2.}
\end{equation}

Two metrics pertaining to the number of output probe photons are of interest: 1) the signal-to-noise ratio (SNR) if the maximum integrated output is desired, or 2) the ratio of the maximum integrated output to the cavity energy if the maximum detection efficiency is desired for a given intensity-dependent dark count rate (dominated by multiphoton absorption, as explored in subsequent sections).

In absence of any non-idealities -- namely multiphoton absorption induced dark counts -- the ability to distinguish the difference signal produced from free carrier dispersion relative to measurement shot noise is of principle interest. If a homodyne measurement is performed on the probe output with a perfectly ``matched" local oscillator field $s_{LO}(t)$ which coherently integrates the difference signal $\delta s_{2-}$, the signal-to-noise ratio (SNR) can be written as
\begin{equation}
\boxed{\text{SNR} = \frac{\mathcal{S}}{\mathcal{N}}=\frac{\int_0^T (2|\delta s_{2-}(t)||s_{LO}(t)|)^2 dt}{|s_{LO}(t)|^2}=4n_p,}
\end{equation}
where the signal $\mathcal{S}$ is the integrated output, and the noise $\mathcal{N}$ is the shot noise of the local oscillator. Therefore, the SNR is an appropriate evaluation metric for the photodetector in absence of any multiphoton absorption. Alternatively, if the objective is to maximize the output SNR while minimizing the intracavity intensity (thereby minimizing the multiphoton absorption-induced dark count rate), the input power-independent expression $n_p/|a|^2$ is an appropriate figure of merit (FOM) which, under the same assumptions implemented to simplify $n_p$, is approximately
\begin{equation}
\boxed{\text{FOM}\equiv \frac{n_p}{|a|^2} \approx \frac{\tau^2\Delta\omega^2\left[ 4\tau e^{-T/\tau}-\tau e^{-2T/\tau}+2T-3\tau\right]}{\tau_2\hbar\omega_0}.}
\label{fom}
\end{equation}

\subsection{Optimization}
\label{sec:optimize}

\begin{figure*}[h!]
\includegraphics[width=.6\textwidth]{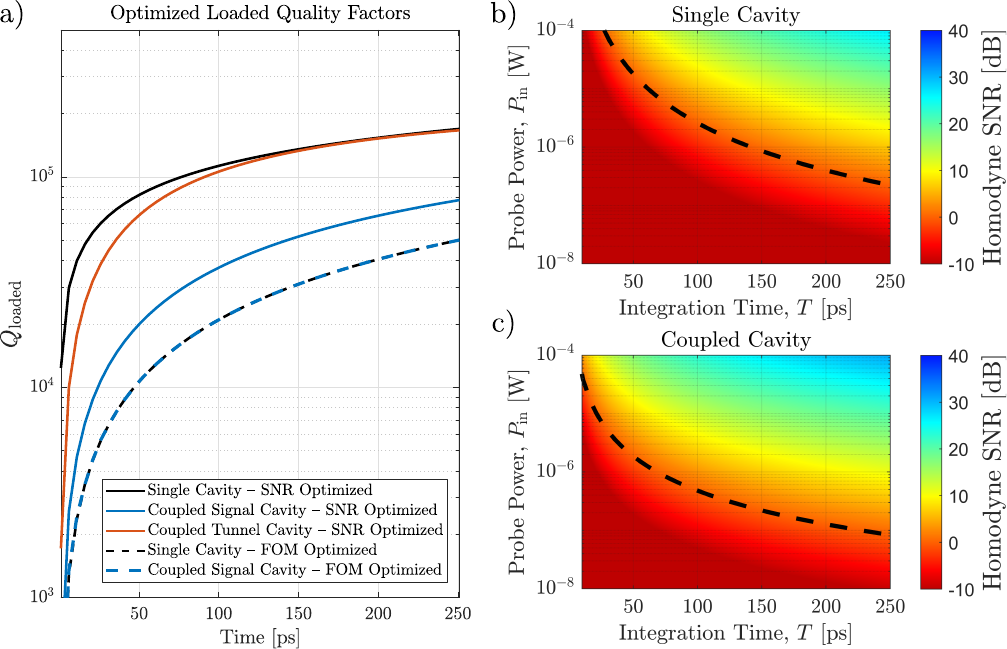}
\caption{Optimized quality factors (a) and homodyne signal-to-noise ratios for the ideal single (b) and coupled (c) cavity detectors. For comparison to Fig.~2 of the main text, each cavity is assumed to have a larger mode volume $V_\text{eff}=10^{-1} (\lambda/n)^3$ at $\lambda=2.3~\upmu$m.}
\centering
\label{compareFig2}
\end{figure*}

These two performance metrics provide independent optimal coupling conditions. Since the functional dependence of $n_p$, and therefore the SNR, upon $\tau_1$ and $\tau_2$ is equivalent, we find a solution which is symmetric in the input and output coupling rates. Therefore, assuming $\tau_1=\tau_2=\tau_c$, we find the optimization criteria
\begin{equation}
\left\lbrace \left[ \frac{1}{k}(\gamma-6)-3\right] \gamma+10\right\rbrace\cosh(\gamma)+\left\lbrace (3\frac{\gamma}{k}-5)(\gamma-1)\right\rbrace\sinh(\gamma)+2\gamma\left(\frac{\gamma}{k}+\frac{3}{k}-1 \right)-10 = 0,
\end{equation}
where $\gamma\equiv T/\tau$ for the loaded decay time $1/\tau = 2/\tau_c+1/\tau_i$, and $k\equiv T/\tau_i$. The result illustrates that the optimum loaded quality factor of the cavity is a function of both the integration time and intrinsic cavity quality factor. The results of this optimization for $Q_i=10^6$,  $V_\text{eff}= 10^{-1}(\lambda/n)^3$, and $\lambda=2.3~\upmu$m --- values representative of a readily fabricable PhC cavity --- are shown in Fig.~\ref{compareFig2} for comparison to the results in the main text (Fig.~2). 

The functional dependence of the FOM upon the coupling rates, on the other hand, is dominated by $\tau_2$. The $\tau_2^{-1}$ term implies that $\tau\approx\tau_2$, as $\tau_2\ll\tau_1,~\tau_i$. Optimizing the figure of merit with respect to $\tau$ yields the optimization constraint
\begin{equation}
(8+4\gamma)e^{-\gamma}-2(1+\gamma)e^{-2\gamma}+2\gamma-6=0,
\end{equation}
which is satisfied for $\gamma\equiv T/\tau\approx 1.9$. Introducing the coupling quality factor of each port, $Q_n=\omega_0/\kappa_n=\omega_0\tau_n/2$, the equivalent condition for coupling $Q$ is 
\begin{equation}
Q_\text{max}^\text{FOM} \approx \frac{\omega T}{3.8}.
\end{equation}
Since $\tau$, and therefore $Q$, is dominated by the fastest leakage path (i.e. the output coupler with decay time $\tau_2$), the optimal solution for the figure of merit exists when $Q_2 \approx Q_\text{max}^\text{FOM}$ and the input coupling is minimized ($Q_1\rightarrow\infty$). The resulting peak FOM is
\begin{equation}
\boxed{\text{FOM}\bigg|_{Q_\text{max}^\text{FOM}} \approx 0.381~\Delta\omega^2 T^2\bigg/\hbar\omega_0.}
\label{FOMsupp}
\end{equation}

When comparing the detection performance and dark count rates, it is convenient to introduce a detection efficiency parameter $\eta_{SN} = 1-e^{-n_p}$, which quantifies the probability of the integrated output signal crossing an optimum threshold level assuming shot noise limited detection \cite{Pratt}. The efficiency at optimal coupling is simply
\begin{equation}
\eta_{SN} = 1-e^{-n_p} = 1-e^{-\text{FOM}|a|^2} \approx 1-e^{-0.381\Delta\omega^2 T^2 |a|^2/\hbar\omega_0},
\label{eta}
\end{equation}
which demonstrates the dependence of the detection efficiency upon the intracavity probe photon number $|a|^2/\hbar\omega_0$.

\subsection{Jitter}
\label{sec:jitter}
We consider photodetector jitter, or timing uncertainty in signal photon detection, originating from three primary sources: 1) fluctuations in the signal photon absorption time, 2) probabilistic variation in the output probe field, and 3) the jitter of the classical homodyne detectors used to measure the phase shift of the transmitted probe field. For visible light, the first can be conservatively estimated by the absorption coefficient of room temperature, undoped silicon ($\alpha_\text{abs}\sim 10^3$ cm$^{-1}$ \cite{Green2008}), which corresponds to a 1/$e$ absorption depth of 10 $\upmu$m and an associated absorption time on the order of 10 fs.

The second contribution, jitter introduced by statistical fluctuations in the probe field photon number, can be estimated using the results of the previous section. Assuming the detector FOM is optimized, Eqn.~\ref{FOMsupp} yields
\begin{equation}
\frac{\Delta t_\text{th}}{t_\text{th}}\approx \frac{1}{2} \frac{\Delta n_p(t_\text{th})}{\langle n_p(t_\text{th}) \rangle},
\label{jitterEst}
\end{equation}
an estimate of the fractional uncertainty in a detection-threshold crossing time $t_\text{th}$ with respect to the mean integrated output photon number $\langle n_p(t_\text{th}) \rangle$. Since the number of probe field photons output from the resonator within a differential time interval is governed by a Poisson distribution, $n_p(t_\text{th})$ at $t_\text{th}$ is similarly Poissonian. For the large photon numbers required for high-efficiency detection, the probability of $m$ photons reaching the detector by $t_\text{th}$ can be approximated by the Gaussian distribution
\begin{equation}
\mathcal{P}(m) \approx \frac{1}{\sqrt{2\pi \langle n_p(t_\text{th})\rangle}}e^{-\frac{\lbrace m-\langle n_p(t_\text{th})\rangle\rbrace^2}{2\langle n_p(t_\text{th}) \rangle}}
\end{equation}
with a mean and variance equal to $\langle n_p(t_\text{th}) \rangle$ \cite{Riley}. If we define the jitter $\Delta t_\text{jitter}$ to be time interval bounded by $\langle n_p(t_\text{th}) \rangle \pm \sigma_{n_p}$, where $\sigma_{n_p}$ is the standard deviation of $\langle n_p(t_\text{th})\rangle$, Eqn.~\ref{jitterEst} simplifies to 
\begin{equation}
\frac{\Delta t_\text{jitter}}{t_\text{th}}\approx  \frac{1}{\sqrt{\langle n_p(t_\text{th}) \rangle}}.
\end{equation}
Assuming $\langle n_p(t_\text{th}) \rangle\gg 1$ as required for efficient photodetection, we see that the jitter from statistical variations of the output field is much smaller than $t_\text{th}$ --- on the order of $10-100$ ps for the fast operation regime of interest --- and can in principle be sub-ps. Therefore, we find that the jitter from the first two sources is negligible in comparison to the jitter associated with the classical probe field detectors.

\section{III. Coupled Cavity Theory}

\begin{figure}[h!]
\includegraphics[width=.45\textwidth]{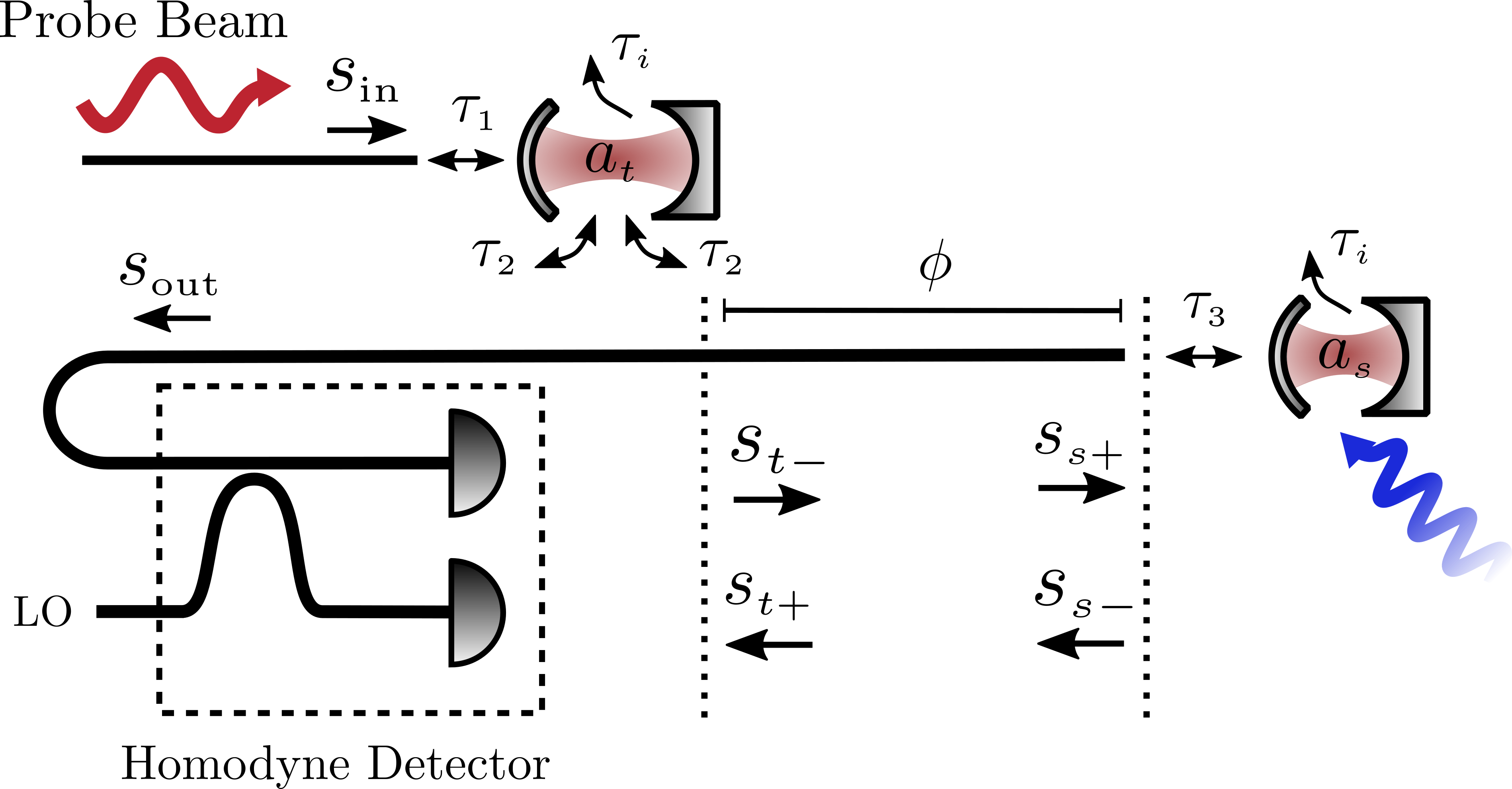}
\caption{Diagram of the coupled cavity photodetector, including signal amplitudes $s$ and coupling rates $1/\tau$. Depending on the spacing $\phi$, destructive interference between the signal cavity ($a_s$) and tunnel cavity ($a_t$) decay paths can enable the signal cavity to be strongly coupled to the output waveguide while maintaining its intrinsic $Q$ in steady state.}
\label{fig: coupledArc}
\end{figure}

The input power required for photodetection can be reduced by implementing a coupled architecture similar to that shown in Figure \ref{fig: coupledArc}. Here, a ``tunnel" cavity is coupled to the output waveguide at a physical distance corresponding to an optical phase shift $\phi$. Destructive interference between $s_{s+}$, the light reflected from the signal cavity towards the output, and the leftward decaying light from the tunnel cavity results in energy storage within the system limited only by the input coupling and intrinsic decay rates. As a direct result of the amplified cavity energies, the required input power is reduced. 

Assuming both cavities are tuned to the resonant frequency $\omega_0$, the system shown in Figure \ref{fig: coupledArc} can be described by the coupled differential equations \cite{Fan2003}
\begin{align}
\frac{da_t}{dt} &= j\omega_0a_t-\underbrace{\left(\frac{1}{\tau_i}+\frac{1}{\tau_1}+\frac{2}{\tau_2}\right)}_{1/\tau_t}a_t + j\sqrt{\frac{2}{\tau_1}}s_\text{in}+j\sqrt{\frac{2}{\tau_2}}s_{t+}\\
\frac{da_s}{dt} &= j\omega_0a_s-\underbrace{\left(\frac{1}{\tau_i}+\frac{1}{\tau_3}\right)}_{1/\tau_s}a_s+j\sqrt{\frac{2}{\tau_3}}s_{s+},
\end{align}
where $1/\tau_1$, $1/\tau_2$, and $1/\tau_3$ are the coupling rates at each port, $1/\tau_i$ is the intrinsic decay rate of the cavities, $1/\tau_s$ ($1/\tau_t$) is the loaded decay rate of the signal (tunnel) cavity, $s_{s+}=j\sqrt{2/\tau_2}e^{j\phi}a_t$, and $s_{t+}=j\sqrt{2/\tau_3}e^{j\phi}a_s+j\sqrt{2/\tau_2}e^{j2\phi}a_t$. For maximum energy storage, we seek to satisfy
\begin{equation}
s_\text{out} = j\sqrt{\frac{2}{\tau_2}}\left(1+e^{j2\phi}\right)a_t^\text{(ss)}+j\sqrt{\frac{2}{\tau_3}}a_s^\text{(ss)}e^{j\phi}=0.
\end{equation}
Inserting the steady state conditions
\begin{align}
a_t^\text{(ss)}&=\frac{j\sqrt{\frac{2}{\tau_1}}s_\text{in}}{j(\omega-\omega_0)+\frac{1}{\tau_t}+\frac{2}{\tau_2}e^{j2\phi}-\frac{4e^{j2\phi}}{\tau_2\tau_3}\left(\frac{1}{j(\omega-\omega_0)+1/\tau_s}\right)}\\
a_s^\text{(ss)}&=\frac{-2e^{j\phi}\sqrt{\frac{1}{\tau_2\tau_3}}}{j(\omega-\omega_0)+\frac{1}{\tau_s}}a_t^\text{(ss)}
\end{align}
and assuming resonant excitation, we find the condition
\begin{equation}
\frac{\tau_3}{\tau_s}\left(1+e^{-j2\phi} \right) = 2.
\end{equation}
To enable rapid extraction of the probe field, the signal cavity is heavily coupled to the output waveguide such that $\tau_3\ll\tau_i$. In this case, $\tau_3\approx\tau_s$, yielding the approximate solution
\begin{equation}
\phi_\text{optimal} = m\pi
\end{equation}
for any integer $m$.

We apply this result to the original system evolution equations, and subsequently configure the coupling rates with a nonlinear optimization routine. The resulting peak integrated output at optimal coupling, as determined by the numerical solution to the system, is shown in Figure 2(d). 

\section{IV. Signal Photon Absorption}
\label{suppAbsSection}

The overall system detection efficiency 
\begin{equation}
\eta=\eta_\text{SN}\eta_{abs}
\end{equation}
is the product of the optimized shot noise-limited detection efficiency and $\eta_{abs}$, the probability of absorbing the signal photon within the probe cavity mode volume. While $\eta_{abs}$ is geometry dependent, we can compute approximate values using a standard L3 geometry ($V_\text{eff}\approx (\lambda/n)^3$; mode profile in Fig.~\ref{figS:L3}) and a subwavelength tip cavity geometry ($V_\text{eff}\approx 10^{-3} (\lambda/n)^3$; mode profile in Fig.~\ref{figS:modeprofile}). The two cavity designs were adapted from \cite{Minkov2014} and \cite{Choi} and scaled to a probe wavelength $\lambda_p = 4.5 ~\upmu$m (corresponding to a slab thickness $t = 0.22~\upmu\text{m}~(4.5~\upmu\text{m}/1.55~\upmu\text{m})\approx 0.64~\upmu\text{m}$). The finite-difference time-domain (FDTD) method was used to simulate the absorption of a wavepacket focused through a thin lens (NA$=1$) onto the surface of each silicon slab geometry. 

\begin{figure*}[h!]
\includegraphics[width=.5\textwidth]{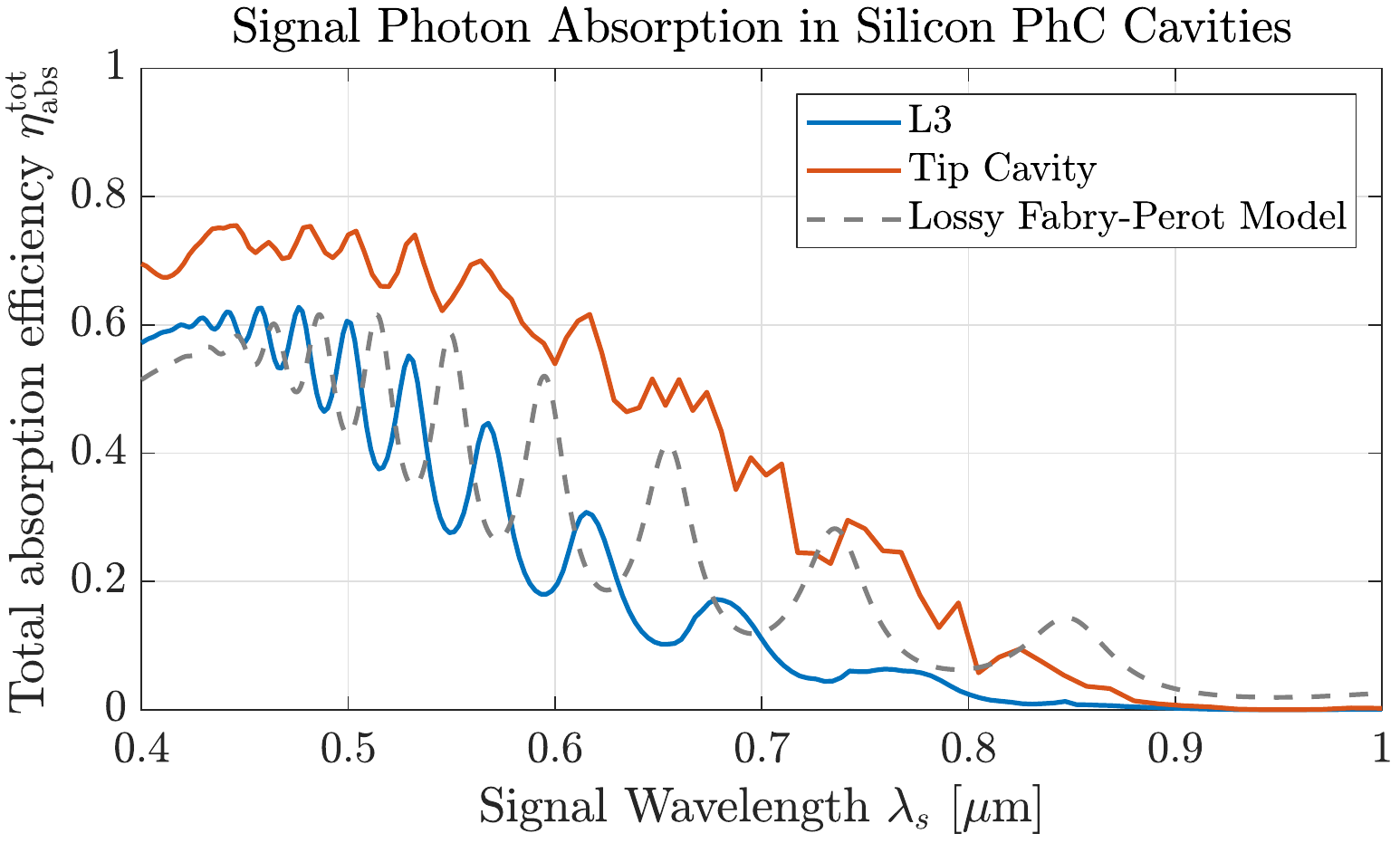}
\caption{Total signal light absorption efficiency $\eta_\text{abs}^\text{tot}$ for standard (L3, $V_\text{eff} \sim (\lambda_p/n)^3$ \cite{Minkov2014}) and subdiffraction mode volume (tip cavity, $V_\text{eff} \sim 10^{-3}~(\lambda_p/n)^3$ \cite{Choi, Hu2018}) silicon photonic crystal cavities at a probe wavelength $\lambda_p = 4.5 ~\upmu$m. At short signal wavelengths, the L3 cavity absorption is well described by a lossy Fabry-Perot resonator model (grey dashed line) \cite{Haus1984}.}
\centering
\label{abs-tot}
\end{figure*}

\begin{figure*}[h!]
\includegraphics[width=\textwidth]{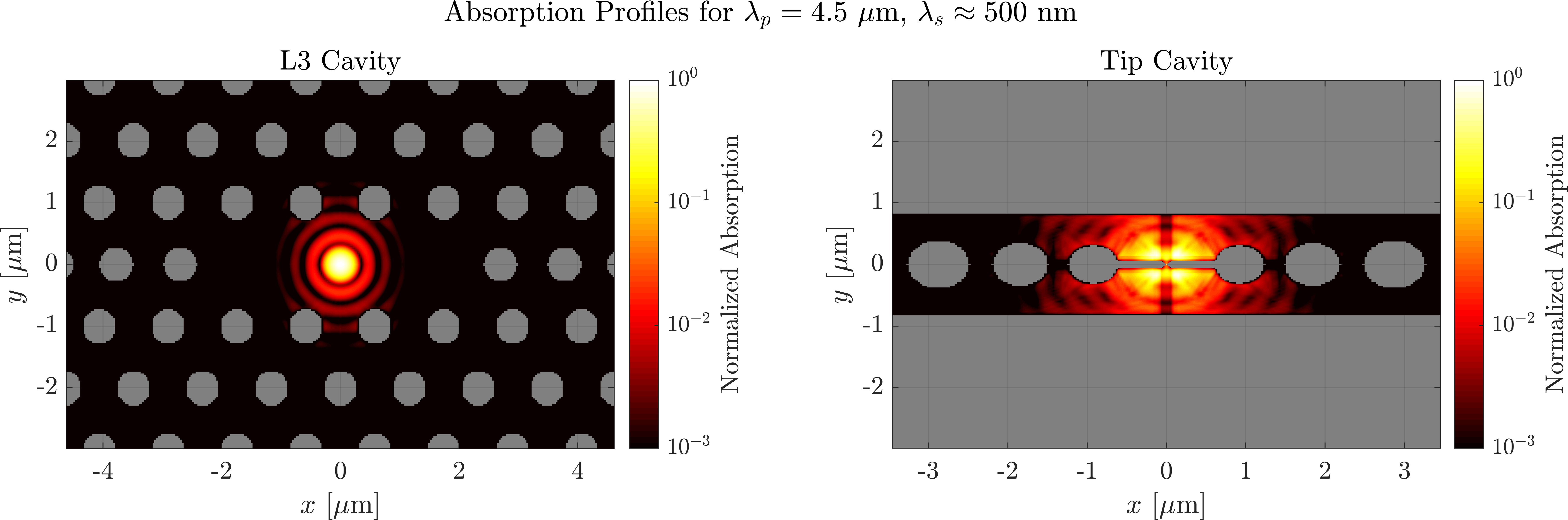}
\caption{Normalized absorption profiles for the L3 and tip cavities for a probe wavelength $\lambda_p = 4.5 ~\upmu$m and a 500 nm signal wavelength.}
\centering
\label{abs-profile}
\end{figure*}

The total absorption $\eta_\text{abs}^\text{tot}$ as a function of signal wavelength $\lambda_s$ for both cavity geometries is shown in Fig.~\ref{abs-tot}. The $\resim 75$\% peak absorption of the tip cavity exceeds that of the L3 geometry ($\resim 60$\%). As a function of wavelength, $\eta_\text{abs}^\text{tot}$ is characterized by a smooth oscillation for the L3 cavity, and by a more complex evolution for the tip cavity. These trends can be understood by considering the absorption profiles in Fig.~\ref{abs-profile} for a visible signal wavelength $\lambda_s = 500$ nm. The results demonstrate that the absorption profile of the L3 cavity lies within the mode volume (compare Fig.~\ref{figS:L3}) and is relatively unaffected by the presence of holes in the PhC slab. Therefore, as shown by the dashed grey line in Fig.~\ref{abs-tot}, the absorption at short wavelengths is well described by a lossy Fabry-Perot resonator model \cite{Haus1984}, where the end mirrors are formed by the two air-silicon interfaces. Alternatively, the absorption profile of the tip cavity extends well beyond the subwavelength feature at the center of the cavity and is significantly modified by the slab geometry.

\begin{figure*}[h!]
\includegraphics[width=\textwidth]{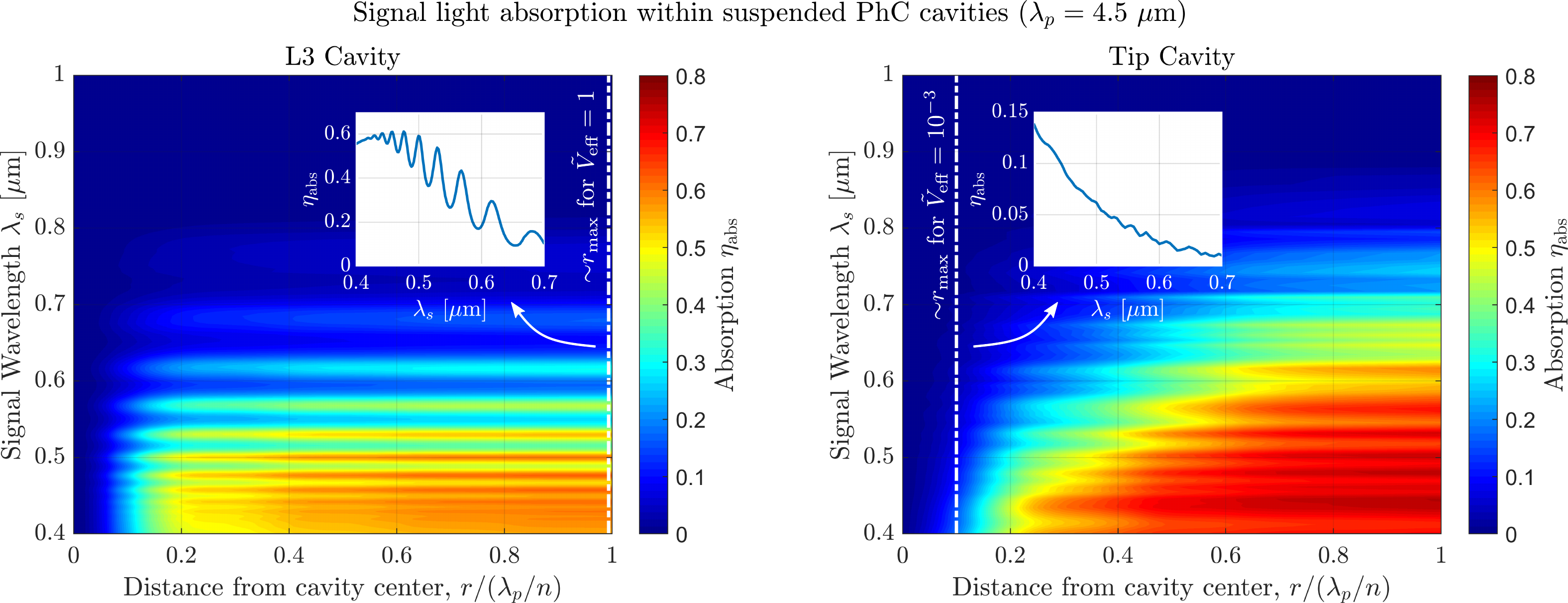}
\caption{Calculated probability of photon absorption $\eta_\text{abs}$ within a radius $r$ from the center of the cavity for the L3 and tip silicon PhC cavities. Dashed vertical lines approximate the maximum radius $r_\text{max} = V_\text{eff}^{1/3}$ for absorption within the mode volume $V_\text{eff}$; insets depict $\eta_\text{abs}$ at $r_\text{max}$.  While the maximum total absorption efficiency $\eta_\text{abs}^\text{tot}$ is larger for the tip cavity (compare Fig.~\ref{abs-tot}), the integrated absorption at $r_\text{max}$ is larger for the L3 cavity due to the larger mode volume.}
\centering
\label{abs-int}
\end{figure*}

Despite the total absorption enhancement afforded by the tip cavity, the absorption efficiency within the probe field mode volume --- $\eta_\text{abs}$, the metric of interest for computing the system detection efficiency $\eta$ --- is larger for the L3 cavity due to its larger mode volume. This is evidenced by Fig.~\ref{abs-int}, which plots the integrated absorption density within a radius $r$ (i.e. the probability of absorbing a signal photon within a radius $r$ from the center of the PhC cavity) for various signal wavelengths. For absorption within the mode volume $V_\text{eff}$, we desire $r\lesssim V_\text{eff}^{1/3}$. The absorption efficiency $\eta_\text{abs}$ at $r_\text{max} = V_\text{eff}^{1/3}$ is plotted in each inset. The results (peak values of $\eta_\text{abs}\sim 0.6$ for the L3 cavity and $\eta_\text{abs}\sim 0.15$ for the tip cavity at visible wavelengths) indicate that $\eta^\text{max}_\text{abs}\approx \eta_\text{abs}^\text{tot}$ for the L3 cavity, but $\eta^\text{max}_\text{abs}\ll \eta_\text{abs}^\text{tot}$ for the tip cavity due to its subwavelength-confined optical mode. 

The incorporation of anti-reflection and reflection coatings on the top and bottom surfaces of the silicon slab, respectively, could improve the absorption efficiency of standard diffraction-limited PhC cavities to $\eta_\text{abs}\sim 1$ at visible wavelengths. For example, perfect transmission and reflection from top and bottom slab surfaces separated by $L \resim 0.5~\upmu$m at $\lambda_s = 400$ nm would yield an absorption efficiency of $\eta_\text{abs}\approx 1-e^{-\alpha 2L} \approx 1-\exp\lbrace 10^5~\text{cm}^{-1}\cdot 1~\upmu\text{m}\rbrace =1-e^{-10}\approx 1$. However, the same technique cannot be applied to the subwavelength cavity, as absorption must be localized to the tip defect region to improve $\eta_\text{abs}$. Further work is therefore required to enhance the localized absorption within ultrasmall mode volume cavities. A few possibilities for future exploration include the design of a doubly-resonant cavity for the probe and signal fields \cite{Rivoire2011, Hueting2014} or the incorporation of a selective absorber, such as a buried heterostructure \cite{Matsuo2010a}, at the center of the cavity.

\section{V. Dark Count Rates and Mode Volumes}
\label{darkCountSection}

\subsection{Dark Count Contributions}

The mean density of thermally-excited free carriers within a semiconductor at a given temperature --- known as the intrinsic carrier concentration $n_i$ --- is governed by the ratio of the material's bandgap to the environmental thermal energy and can be approximated with a Fermi-Dirac distribution. Due to the relatively large bandgap of silicon ($E_g$=1.1 eV), this relationship yields a negligible intra-cavity thermal carrier concentration for our proposed system. Specifically, for silicon at 300 K, the intrinsic carrier concentration is approximately $1.5\times 10^{10}~\text{cm}^{-3}$, which leads to mean electron-hole pair population $\bar{n}_\text{cavity}$ of $\resim 4\times10^{-6}$ within the proposed cavity volume ($V_\text{eff}= 10^{-3}(\lambda/n_\text{Si})^3$) at any given time. Since the mean free carrier lifetime ($\resim$ns \cite{Tanabe2008}) is orders of magnitude longer than the detector integration times of interest ($\resim 10-100$ ps), we assume a quasi-static carrier distribution during the detector integration time. In this case, the probability of a non-zero thermally-induced free carrier concentration can be estimated as
\begin{equation}
\mathcal{P}\lbrace n_\text{cav}> 0\rbrace=1-\mathcal{P}\lbrace n_\text{cav}= 0\rbrace=1-\frac{1}{\bar{n}_\text{cavity}+1}\left(\frac{\bar{n}_\text{cavity}}{\bar{n}_\text{cavity}+1}\right)^{n_\text{cavity}}\bigg|_{n_\text{cavity}=0}=1-\frac{1}{1+\bar{n}_\text{cavity}}\approx \bar{n}_\text{cavity}.
\end{equation}
We therefore see that at 300 K the dark count probability is on the order of $4\times10^{-6}$. Additionally, modest cooling dramatically reduces the thermal excitation probability. Since $n_i \propto T_0^{3/2}e^{-E_g/2k_BT_0}$, where $k_B$ is the Boltzmann constant and $T_0$ is the ambient temperature, cooling silicon to 77 K decreases the intrinsic carrier concentration to on the order of $10^{-18} \text{ cm}^{-3}$. Given these metrics, we find that the density of thermally excited free-carriers is negligible. 

Temperature \textit{fluctuations}, on the other hand, cannot be ignored if the temperature dependence of the cavity material's refractive index leads to index changes on the order of those produced by free carrier dispersion. The cavity therefore requires a temperature stability 
\begin{equation}
\Delta T_0 \leq \Delta n_\text{Si}/\alpha_\text{TO}
\end{equation}
which is dictated by the thermo-optic coefficient $\alpha_\text{TO}$ of the cavity material. For our proposed cavity, silicon's thermo-optic coefficient of $\resim 1.8\times10^{-4}$ \cite{Xu} thus demands sub-Kelvin ($\resim 0.1$K) temperature stability. The integration of a modern PID temperature controller, which can achieve sub-millikelvin temperature stabilization \cite{Lee2016}, will therefore minimize the rate of dark counts generated by temperature variations of the substrate.

\begin{figure}[h!]
\includegraphics[width=.6\textwidth]{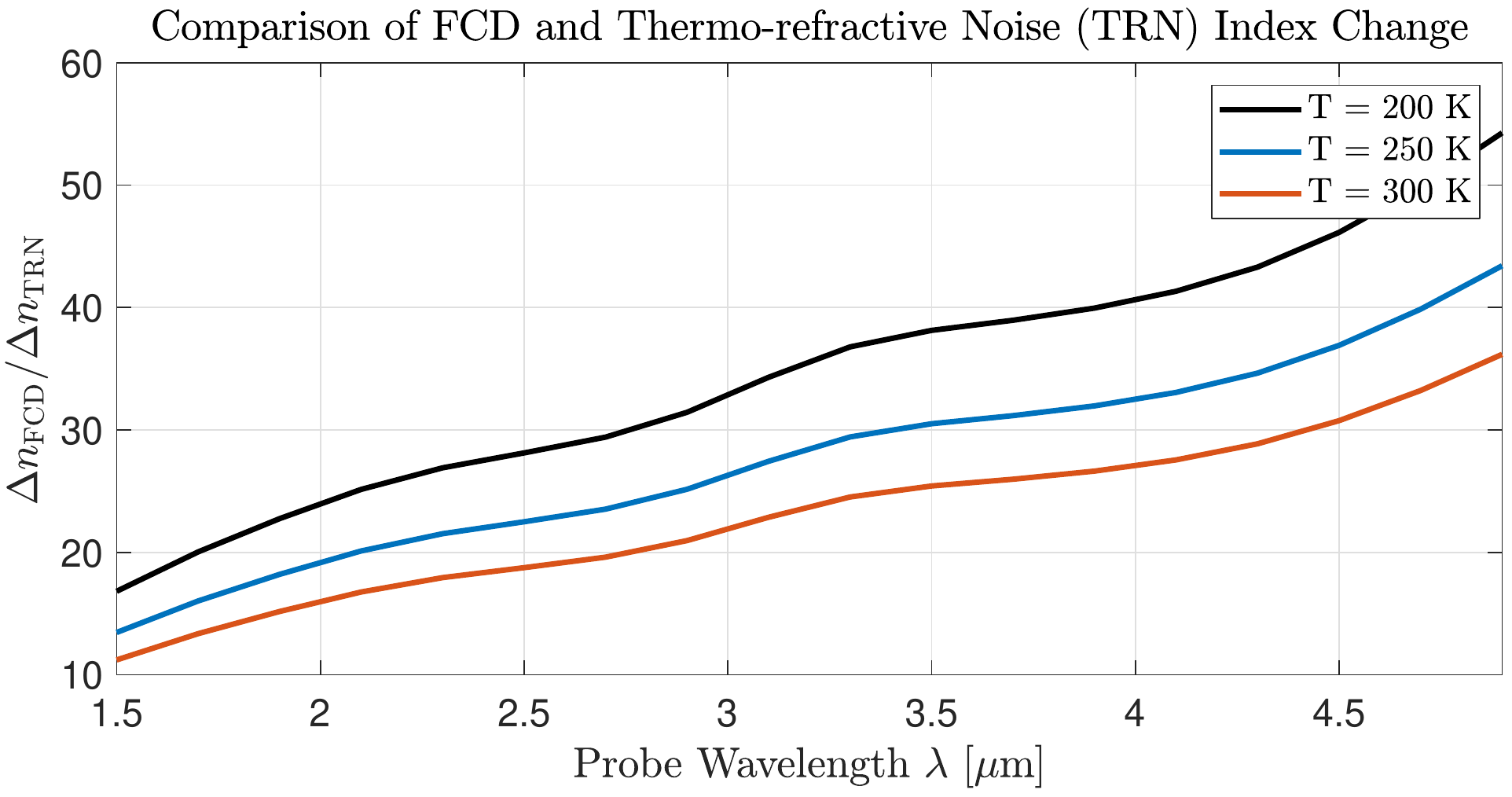}
\caption{Comparison of relative magnitude of single photon-induced free carrier dispersion and thermo-refractive noise index changes for various probe wavelengths and temperatures assuming the parameters stated in the text.}
\label{fig: TRN}
\end{figure}

While slowly-varying (relative to the detection interval) temperature deviations can be high-pass filtered, fundamental statistical temperature fluctuations must still be considered. Within a semiconductor nanocavity of heat capacity $C_V$ at temperature $T$, the magnitude of these fluctuations can be estimated as \cite{Gorodetsky2004,Saurav}
\begin{equation}
\delta T \approx \sqrt{\frac{k_B T^2}{\rho C_V V_T}},
\end{equation}
where $k_B$ is the Boltzmann constant and $V_T = V_\text{eff}^2/V_\text{eff}^{(2)}$ for the second order effective mode volume $V_\text{eff}^{(2)}$ (as defined in Eqn. \ref{effModeVolumes}) assuming complete confinement within the dielectric medium. Given $T=300$ K, $\rho=2.3$ g/cm$^3$, $C_V=0.7$ J/g$\cdot$K, $V_\text{eff}=10^{-3}(\lambda/n_\text{Si})^3$, and $V_\text{eff}^{(2)}\approx 0.075 V_\text{eff}$ (as estimated from the ``tip" cavity profile in Fig. \ref{figS:modeprofile}), Fig. \ref{fig: TRN} shows that the resulting temperature fluctuations yield index changes that are an order of magnitude weaker than those produced from the absorption of a single photon. Therefore, we omit thermo-refractive noise from our analysis.

\subsection{Multiphoton Absorption Dark Count Rates}
\label{sec:darkcounttheory}
As discussed in the main text, the dark count rate is dominated by multiphoton absorption events due to the large cavity energies required for sufficient detection efficiency. The dark count rate from a $k$th order nonlinear absorption ($k$PA) rate $\frac{dn^k_\text{MPA}}{dt}$ (under the assumption that each multiphoton absorption event contributes a dark count, as the output generated from the event is presumed to be indistinguishable from that created by the absorption of a signal photon) is \cite{Nathan1985}
\begin{equation}
R_\text{dark}=\frac{dn^{(k)}_\text{MPA}}{dt} = \frac{\beta_k}{k\hbar\omega_0}I_\text{max}^k V_{k\text{PA}},
\end{equation}
where $V_{k\text{PA}}$ is the mode volume for $k$-photon absorption processes (derived in the subsequent section), $\beta_k$ is the corresponding multiphoton absorption coefficient, and $I_\text{max}= |a|^2c/2n_\text{Si}V_\text{eff}$ is the peak intracavity intensity for a silicon cavity with index $n_\text{Si}$, optical mode volume $V_\text{eff}$, and stored energy $|a|^2$. The dark count rate can thus be approximated as
\begin{equation}
R_\text{dark} = \frac{\beta_k}{k\hbar\omega_0}\left(\frac{c}{2n_\text{Si}}\right)^k\frac{V_{k\text{PA}}}{V_\text{eff}^k}|a|^{2k}
\end{equation}
We can instead parameterize $R_\text{dark}$ as a function of the detection efficiency using Eqn. \ref{eta}, yielding
\begin{equation}
R_\text{dark} = \frac{(\hbar\omega_0)^{k-1}\beta_k}{k(0.381\Delta\omega^2T^2)^k}\left(\frac{c}{2n_\text{Si}}\right)^k\frac{V_{k\text{PA}}}{V_\text{eff}^k}\ln\left(\frac{1}{1-\eta_{SN}}\right)^k,
\label{dcr}
\end{equation}
which, due to the wavelength dependence of the photon energy ($\hbar\omega_0$), resonant frequency perturbation, the most prominent $k$-photon absorption process, the material index ($n_\text{Si}$), and mode volume ($V_\text{eff}= \zeta(\lambda/n_\text{Si})^3$ for a given ``mode volume fraction" $\zeta$), is itself highly wavelength dependent.

The resulting expression in Equation \ref{dcr} suggests that, as expected, both the optical and higher-order multiphoton absorption mode volumes should be minimized in order to achieve a smaller dark count rate. In standard photonic crystal cavities, the intensity variation within the optical mode volume is relatively small, such that the multiphoton absorption mode volume can be approximated as equal to that of the optical mode (for example, $V_\text{2PA}=2\sqrt{2}V_\text{eff}$ assuming a Gaussian intensity distribution in a uniform medium). However, the ultrasmall mode volume ($V_\text{eff}\ll (\lambda/n)^3$) photonic crystal cavities of interest for this work \cite{Hu2016,Choi,Hu2018} contain localized defects which amplify the optical intensity within small regions of the diffraction-limited mode size, $V_\text{min} \sim (\lambda/2n)^3$ \cite{Boroditsky1998}. In this case, as further explored in the following section, $V_{k\text{PA}}$ -- and therefore the dark count rate for a given detection efficiency -- can be several orders of magnitude lower than the value provided by this estimate.

\subsection{Dark Count Mode Volumes -- $V_{k\text{PA}}$}
As mentioned in the previous section, the ``tip" type cavities presented in \cite{Hu2016,Choi,Hu2018} leverage boundary conditions at the semiconductor-air interfaces in order to generate a localized, high intensity region within the diffraction limited mode volume $V_\text{min} \sim (\lambda/2n)^3$. The corresponding reduction in optical mode volume $V_\text{eff}$ can be deduced from the common definition,
\begin{equation}
V_\text{eff} = \frac{\int \epsilon(\vec{r})|\vec{E}(\vec{r})|^{2} d^3\vec{r}}{\text{max}\left\lbrace \epsilon|\vec{E}|^{2} \right\rbrace},
\end{equation}
which can be viewed as an energy density-based weighted average of volume. If the high intensity region created by a single cavity defect (a tip, for example) is sufficiently well-localized, the integrated energy density over the optical mode is roughly constant, while the peak energy density $\text{max}\left\lbrace \epsilon |\vec{E}|^{2} \right\rbrace$ increases substantially. The optical mode volume is therefore reduced by an amount roughly proportional to the increase in peak energy density induced through the introduction of the defect structure. The resulting field inhomogeneity also complicates the behavior of higher order mode volumes, which can be defined as
\begin{equation}
V_\text{eff}^{(k)} = \frac{\int_\text{mode}  \epsilon(\vec{r})^k|\vec{E}(\vec{r})|^{2k} d^3\vec{r}}{\text{max}\left\lbrace \epsilon^k|\vec{E}|^{2k} \right\rbrace},
\label{effModeVolumes}
\end{equation}
for $k>1$. For a typical fundamental photonic crystal cavity mode where the field is relatively uniform over the mode volume $V_\text{eff}$, $V_\text{eff}^{(k>1)}$ should be approximately on the same order of magnitude as $V_\text{eff}$. On the other hand, for tip-based cavities, higher order mode volumes weigh the high intensity region more heavily, as the integral of the normalized energy density outside of the defect region is significantly reduced while that within the small, high intensity region is only marginally affected. The scaling of the higher order mode volumes for these defect-based cavities is therefore dependent upon both the peak intensity within the defect region, as well as the volume of this region. Since the multiphoton absorption (MPA) rate is proportional to $I^k$ for $k\geq  2$, we are interested in defining an ``effective" MPA volume, and analyzing its scaling as a function of absorption event order. 

The dark count rate density -- the number of multiphoton absorption events per unit volume per unit time -- for $k$th order absorption events at a position $\vec{r}$ is defined as \cite{Nathan1985}
\begin{equation}
r_\text{dark}(\vec{r}) = \frac{\beta_k}{k\hbar\omega_0}I(\vec{r})^k.
\end{equation}
For a non-uniform intensity profile, the total dark count rate can be directly evaluated by integrating the dark count rate density over the semiconductor, where MPA can occur:
\begin{equation}
\label{eq:dcrInt}
R_\text{dark} = \int_\text{semi} r_\text{dark}(\vec{r}) d^3\vec{r} = \int_\text{semi} \frac{\beta_k}{k\hbar\omega_0}I(\vec{r})^k d^3\vec{r}.
\end{equation}
Alternatively, the integration can be replaced through the introduction of an ``effective MPA mode volume" $V_{k\text{PA}}$, which leads to the simple relation
\begin{equation}
R_\text{dark} = \frac{\beta_k}{k\hbar\omega_0}I_\text{max}^k V_{k\text{PA}}
\end{equation}
where $I_\text{max} = |a|^2c/n V_\text{eff}$ is the peak intensity resulting from the confinement of a total cavity energy $|a|^2$  within the standard optical mode volume $V_\text{eff}$. Here, we seek to define $V_{k\text{PA}}$. Given $I(\vec{r})= (c/2n)\epsilon_0|\vec{E}|^2$, Equation \ref{eq:dcrInt} simplifies to 
\begin{equation}
R_\text{dark} = \frac{\beta_k}{k\hbar\omega_0}\int_\text{semi} \left(\frac{c}{2n(\vec{r})}\right)^k \epsilon(\vec{r})^k|\vec{E}(\vec{r})|^{2k} d^3\vec{r}.
\end{equation}
Assuming a homogeneous semiconductor medium of index $n_\text{semi}$, we find
\begin{equation}
R_\text{dark} = \frac{\beta_k}{k\hbar\omega_0}\left(\frac{c}{2n_\text{semi}}\right)^k \int_\text{semi}  \epsilon_\text{semi}^k|\vec{E}(\vec{r})|^{2k} d^3\vec{r},
\end{equation}
which can be re-expressed as follows to realize the standard mode volume form:
\begin{equation}
R_\text{dark} = \frac{\beta_k}{k\hbar\omega_0}\underbrace{\left(\frac{c}{2n_\text{semi}}\text{max}\left\lbrace \epsilon_\text{semi}|\vec{E}|^2 \right\rbrace\right)^k}_{I_\text{max}^k} \underbrace{\frac{\int_\text{semi}  \epsilon_\text{semi}^k|\vec{E}(\vec{r})|^{2k}d^3\vec{r}}{\text{max}\left\lbrace \epsilon_\text{semi}^k|\vec{E}|^{2k} \right\rbrace}}_{V_{k\text{PA}}}.
\label{eq:definingMPAvol}
\end{equation}
Therefore, the desired MPA mode volume is
\begin{equation}
\boxed{V_{k\text{PA}}\equiv \frac{\int_\text{semi}  \epsilon_\text{semi}^k|\vec{E}(\vec{r})|^{2k}d^3\vec{r}}{\text{max}\left\lbrace \epsilon_\text{semi}^k|\vec{E}|^{2k} \right\rbrace} = \frac{\int_\text{semi}  |\vec{E}(\vec{r})|^{2k}d^3\vec{r}}{\text{max}\left\lbrace |\vec{E}|^{2k} \right\rbrace}.}
\end{equation}
Intuitively, $V_{k\text{PA}}$ is analogous to the standard $V_\text{eff}$ with two differences: 1) it is evaluated as a weighted average of volume with respect to higher orders of the field energy density as desired for MPA processes ($\propto I^k$), and 2) it is only evaluated over the dielectric region of the PhC, where MPA can occur.

To verify the consistency of the alternative definition of $I_\text{max}$ presented in Equation \ref{eq:definingMPAvol}, we can expand the originally provided definition $I_\text{max} = |a|^2c/2n_\text{semi}V_\text{eff}$:
\begin{equation}
I_\text{max} = \frac{c}{2n_\text{semi}}\underbrace{\left( \frac{|a|^2}{V_\text{eff}} \right)}_{U} = \frac{c}{2n_\text{semi}} \left( \frac{\int_\text{mode}  \epsilon(\vec{r})|\vec{E}(\vec{r})|^{2} d^3\vec{r}}{\frac{\int_\text{mode}  \epsilon(\vec{r})|\vec{E}(\vec{r})|^{2} d^3\vec{r}}{\text{max}\left\lbrace \epsilon|\vec{E}|^{2} \right\rbrace}} \right) = \frac{c}{n_\text{semi}}\text{max}\left\lbrace \epsilon|\vec{E}|^2 \right\rbrace.
\end{equation}

Using simulated mode profiles of defect-based ultrasmall mode volume cavities, we can calculate thee newly defined MPA mode volume to determine its scaling with respect to $k$. Ideally, we desire a cavity design in which $V_{k\text{PA}}\ll V_\text{eff}$, such that the net dark count rate is reduced from that provided by a photonic crystal cavity where  $V_{k\text{PA}}\sim V_\text{eff}$.

\begin{figure*}[h!]
\includegraphics[width=.9\textwidth]{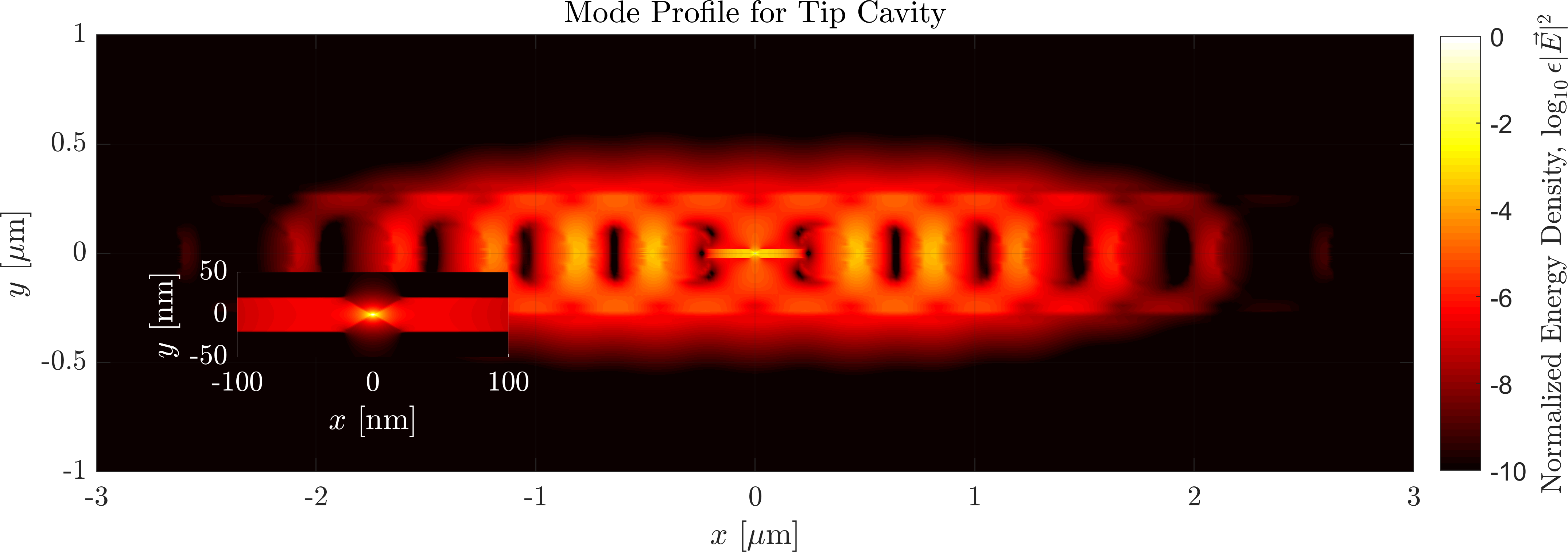}
\caption{Energy density profile at the central plane within the tip-defect silicon cavity described in \cite{Choi} for ultrastrong probe confinement ($\lambda = 1550$ nm). The inset shows a zoomed view of the tip at the center of the cavity (note that a different color scale is use for visualization). Notably, the high intensity region at the tip is localized to a region much smaller than the diffraction limited mode volume.}
\centering
\label{figS:modeprofile}
\end{figure*}

Given this goal, we considered the tip-based silicon cavity presented in \cite{Choi}. The mode profile, shown in Figure \ref{figS:modeprofile}, is characterized by a single, well-localized high intensity region at the center of the silicon tip. Since the effective mode volume is inversely proportional to the peak intensity, the presence of this region significantly reduces the effective mode volume to the sub-wavelength regime ($V_\text{eff}\sim 10^{-5} (\lambda/n_\text{Si})^3$).

\begin{figure*}[h!]
\includegraphics[width=.6\textwidth]{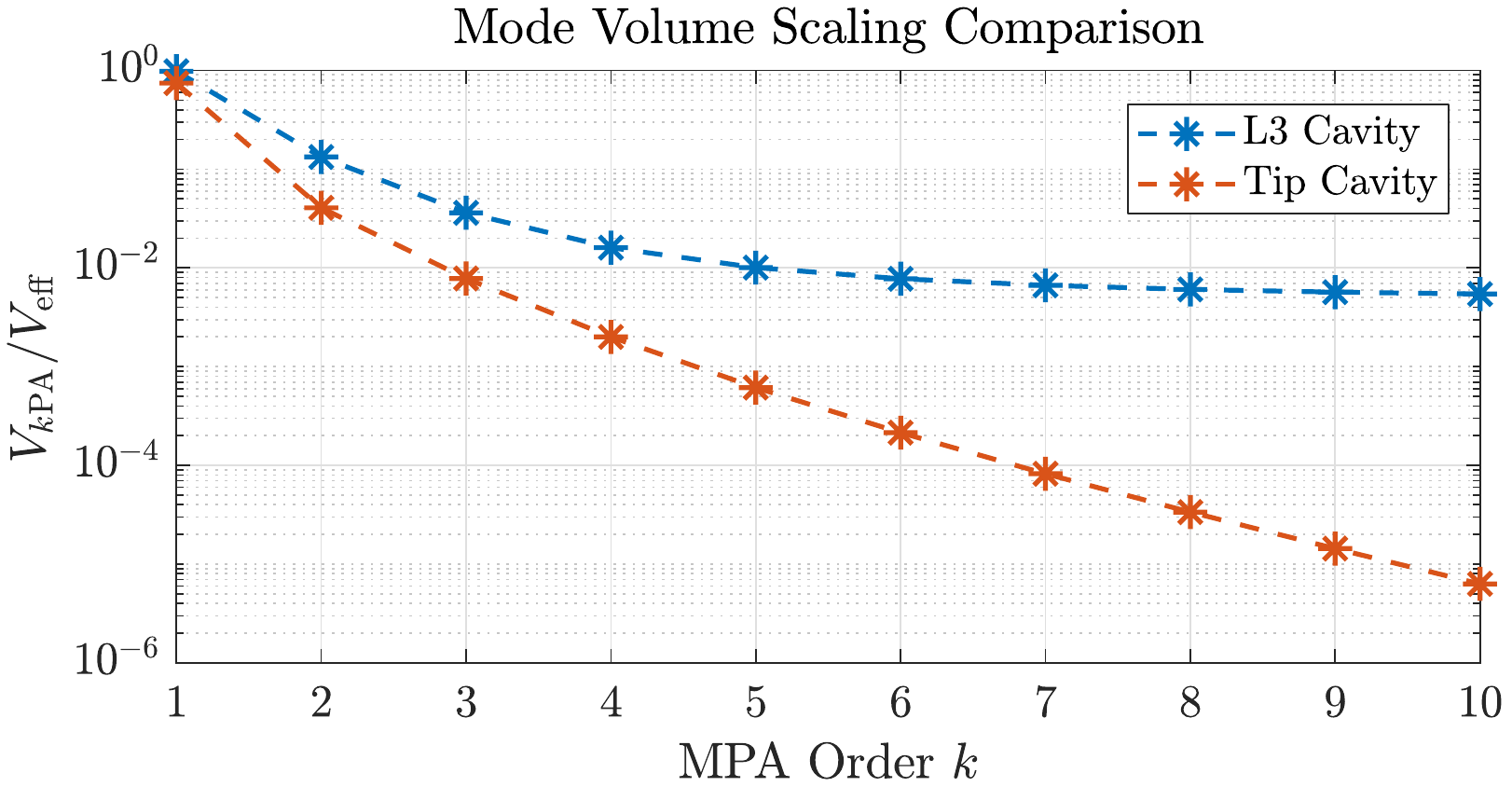}
\caption{Scaling of multiphoton absorption volumes $V_{k\text{PA}}$ for higher order processes in standard (represented here by an L3 cavity with the 2D mode profile shown in Fig.~\ref{figS:L3}) and sub-wavelength confining ``tip" PhC cavities (see Fig.~\ref{figS:modeprofile}).}
\centering
\label{figS:volScaling}
\end{figure*}

\begin{figure*}[h!]
\includegraphics[width=.6\textwidth]{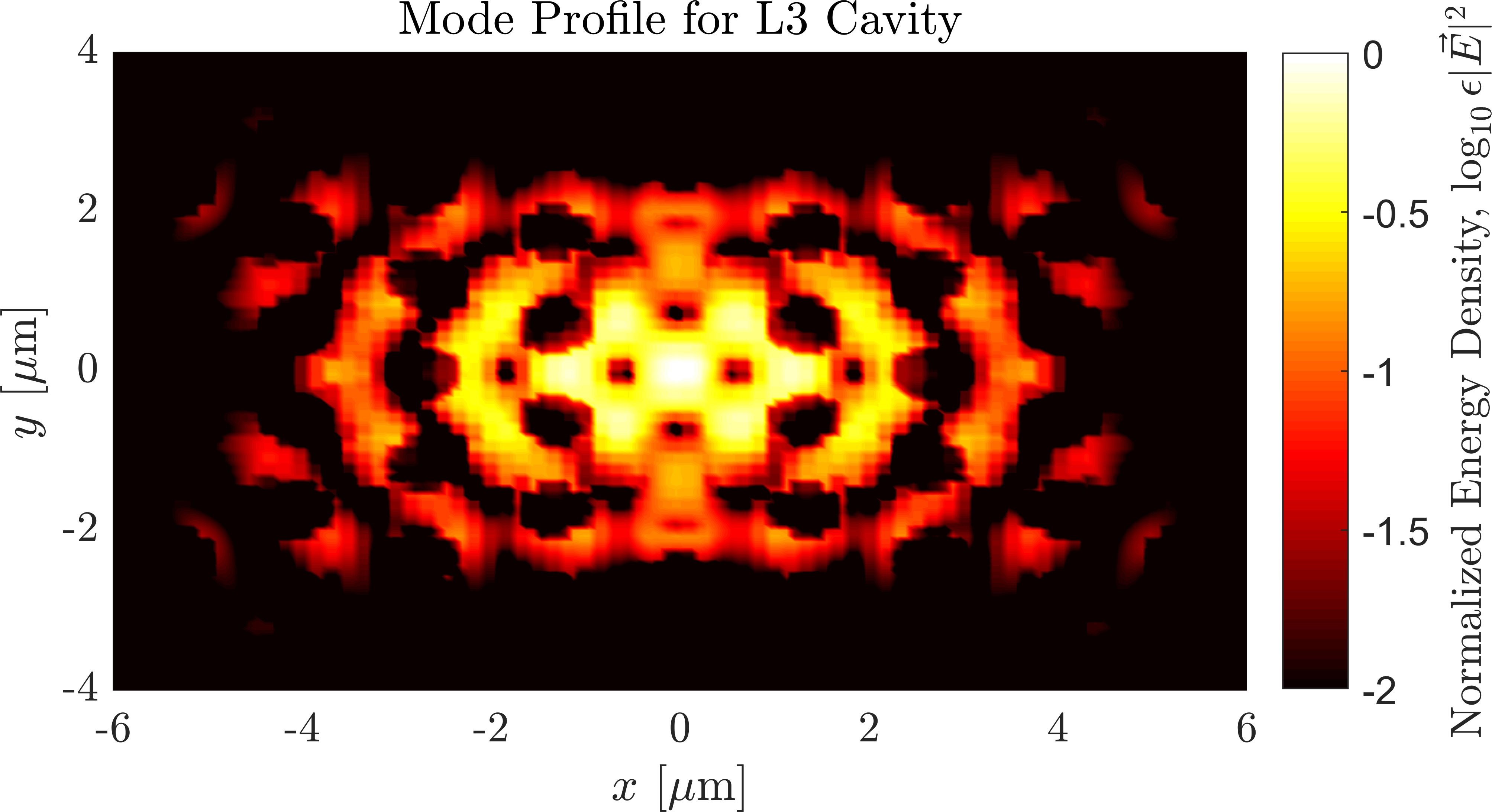}
\caption{Mode profile of the L3 cavity used for the mode volume comparison in Fig.~\ref{figS:volScaling}.}
\centering
\label{figS:L3}
\end{figure*}

The derived multiphoton absorption volume was evaluated for this field profile, yielding the scaling shown in Figure \ref{figS:volScaling}. At higher orders, the confinement within silicon weakens, leading to a continual decay of the MPA mode volume. For the MPA orders of interest (3PA, 4PA, and 5PA, corresponding to probe wavelengths between 2.3 and 4.5 microns), $V_{k\text{PA}}/V_\text{eff}$, decreases from $\resim 10^{-2}$ to $\resim 10^{-4}$. Comparing this scaling to that of a standard L3 photonic crystal cavity (with $V_\text{eff}\sim 0.5 (\lambda/n_\text{Si})^3$ and a mode profile shown in Fig.~\ref{figS:L3}) reveals a second advantage of the defect-based cavities (in addition to ultrasmall probe confinement) -- suppression of higher order noise processes.

\section{VI. Probe Squeezing Enhancement}
\begin{figure}
\includegraphics[width=.5\textwidth]{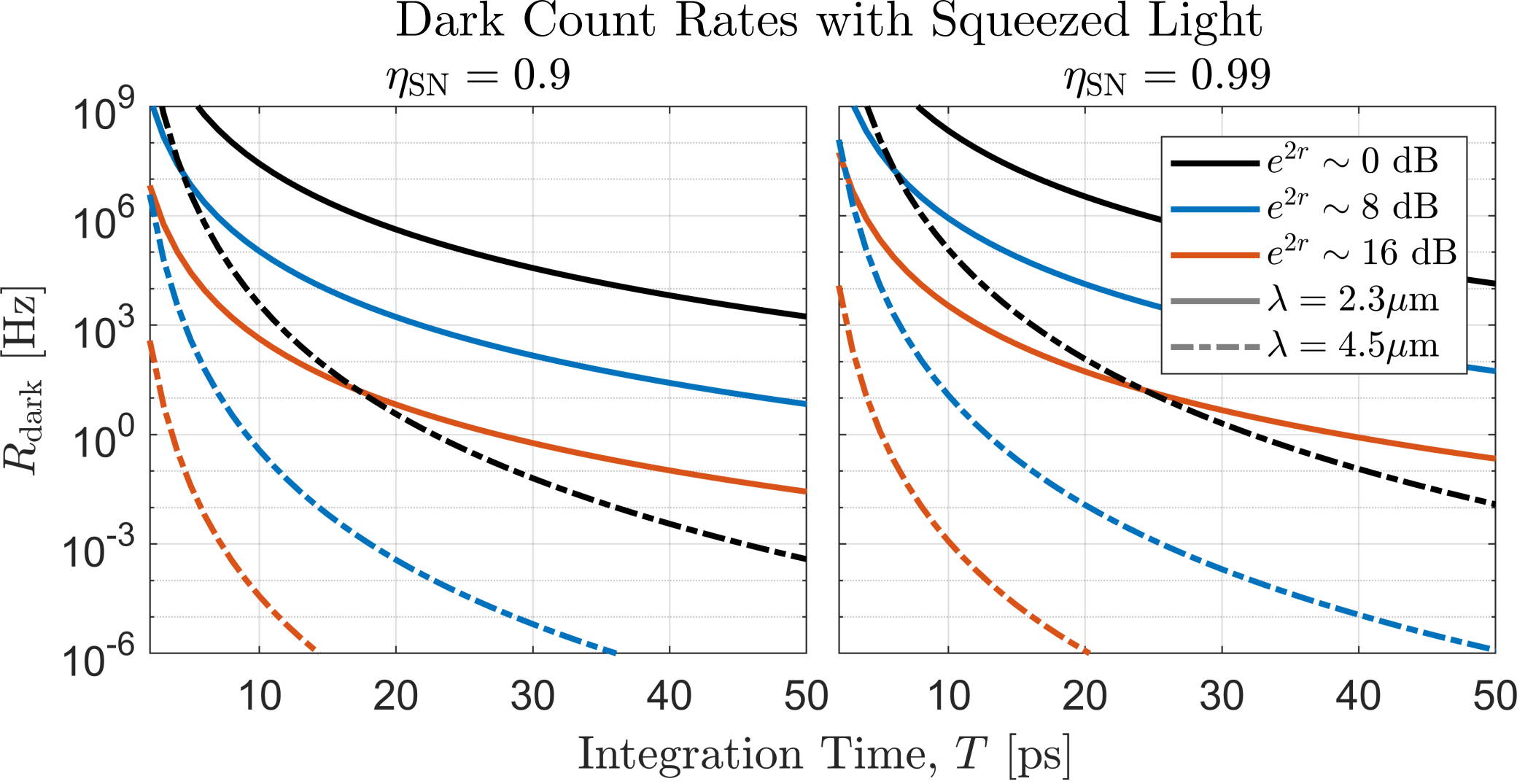}
\caption{Effects of probe squeezing on dark count rates for two detection efficiencies of interest at a probe wavelength of 2.3 $\upmu$m (solid) and 4.5 $\upmu$m (dashed).}
\centering
\label{fig: squeeze}
\end{figure}

As shown in Fig.~\ref{fig: squeeze}, a phase-squeezed probe beam could be used to even further suppress these rates. Since $R_\text{dark}\propto \text{FOM}_\text{opt}^{-k}$, the $e^{2r}$ factor of SNR improvement afforded by a squeezed state with squeezing parameter $r$ decreases the dark count rate by a factor of $e^{2kr}$ \cite{Loudon2000}. This improvement is therefore especially significant for higher orders of MPA. For example, 10 dB of squeezing reduces $R_\text{dark}$ by a factor of $10^3$ at 2.3 $\upmu$m, and by a factor of $10^5$ at 4.5 $\upmu$m.

%